\documentclass[aps,prd,superscriptaddress,showpacs,showkeys,a4paper,superscriptaddress,10pt,notitlepage,twocolumn]{revtex4-1}

\usepackage[utf8]{inputenc}
\usepackage[english]{babel}
\usepackage{amsmath}
\usepackage{amssymb}
\usepackage{amsfonts}
\usepackage{graphicx}
\usepackage{bm}
\usepackage{cancel}         
\usepackage{mathtools}             
\usepackage[T1]{fontenc}
\usepackage{comment}
\usepackage{subfigure}
\usepackage{longtable}
\usepackage{soul}
\usepackage[dvipsnames]{xcolor}

\usepackage{xcolor}
\definecolor{link_blue}{RGB}{52,46,157}

\usepackage[colorlinks,citecolor=link_blue,linkcolor=link_blue,urlcolor=link_blue]{hyperref}
\usepackage[tiny,raggedright]{titlesec}             
\usepackage[a4paper,textwidth=480pt,textheight=700pt,top=80pt,left=60pt]{geometry}

\renewcommand{\vec}{\boldsymbol}

\newcommand\ri{\mathrm{i}}

\DeclareMathOperator{\re}{Re}
\DeclareMathOperator{\im}{Im}

\allowdisplaybreaks[2]

\newcommand{\change}{\textcolor{black}}

\def\no#1{{\color{black}{#1}}}  
\def\CHK#1{{\color{black}{#1}}}

\begin{document}

\title{Relativistic effective charge model of a multi-electron atom}

\author{K. Dzikowski}

\affiliation{Max Planck Institute for Nuclear
  Physics, Saupfercheckweg 1, 69117 Heidelberg, Germany}

\author{O. D. Skoromnik}
\email[Corresponding author: ]{olegskor@gmail.com}                

\affiliation{Max Planck Institute for Nuclear Physics, Saupfercheckweg
  1, 69117 Heidelberg, Germany}

\author{I. D. Feranchuk}

\affiliation{Atomic Molecular and Optical Physics Research Group,
  Advanced Institute of Materials Science, Ton Duc Thang University,
  19 Nguyen Huu Tho Str., Tan Phong Ward, District 7, Ho Chi Minh
  City, Vietnam}

\affiliation{Faculty of Applied Sciences, Ton Duc Thang University, 19
  Nguyen Huu Tho Str., Tan Phong Ward, District 7, Ho Chi Minh City,
  Vietnam}


\author{N. S. Oreshkina}

\affiliation{Max Planck Institute for Nuclear Physics, Saupfercheckweg
  1, 69117 Heidelberg, Germany}

\author{C. H. Keitel}

\affiliation{Max Planck Institute for Nuclear Physics, Saupfercheckweg
  1, 69117 Heidelberg, Germany}

\begin{abstract}
  A relativistic version of the effective charge model for computation
  of observable characteristics of multi-electron atoms and ions is
  developed. A complete and orthogonal Dirac hydrogen basis set,
  \no{depending} on one parameter --- effective nuclear charge $Z^{*}$
  --- identical for all single-electron wave functions of a given atom
  or ion, is employed for the construction of the secondary-quantized
  representation. The effective charge is uniquely determined by the
  charge of the nucleus and a set of \no{electron} occupation numbers
  for a given state. We thoroughly study the accuracy of the
  \no{leading-order approximation for the total binding energy} and
  demonstrate that it is independent of the number of electrons of a
  multi-electron atom. In addition, it is shown that the fully
  analytical \no{leading-order} approximation is especially suited for
  the description of highly charged ions since our wave functions are
  almost coincident with the Dirac-Hartree-Fock ones for the complete
  spectrum. Finally, \CHK{we evaluate various} \change{atomic
    characteristics, such as scattering factors and photoionization
    cross-sections, and thus} \CHK{envisage} \change{that the
    effective charge model can replace other models of comparable
    complexity, such as} the Thomas-Fermi-Dirac model for all
  applications where it is still utilized.
\end{abstract}



\maketitle

\section{Introduction}
\label{sec:introduction}

Despite great advances in the accuracy and efficiency of modern
numerical methods \cite{filippin_multiconfiguration_2016,
  puchalski_relativistic_2017, peach_fractional_2015,
  jensen2007introduction}, simple analytical models for describing
multi-electron atoms and ions are still actively developed
\cite{gomez_simple_2019, tatewaki_relativistic_2018,
  karpov_atomic-number_2017, hey_forms_2017}. The most popular of them
include the multi-parametric screening hydrogen orbitals
\cite{hau2012high, seaton_quantum_1983}, Slater orbitals
\cite{slater_atomic_1930} and the semi-classical Thomas-Fermi (TF)
model \cite{thomas_1927, landau1981quantum}. These models are widely
used, whenever certain level of precision has to be sacrificed for
improved computation time, e.g. in computational plasma physics
\cite{ciricosta_direct_2012, starrett_thomas-fermi_2017,
  lee_model_1987, scott_cretinradiative_2001, chung_flychk:_2005,
  dyachkov_region_2016}, semiconductors
\cite{schulze-halberg_position-dependent_2013, smith_electronic_2014},
screening effects in the cross sections of bremsstrahlung and pair
production \cite{tsai_pair_1974, olsen_photon_1959,
  davies_theory_1954, seltzer_bremsstrahlung_1985,
  poskus_brems:_2018}, X-ray scattering and diffraction
\cite{hau2012high, feranchuk_new_2002}, and crystallography
\cite{toraya_new_2016}.

In addition, the wave functions obtained from these simple analytical
models are used as the initial approximation for the solution of
Hartree-Fock (HF) \cite{FockA1930naeherungs, fischer1977hartree,
  grant2007relativistic, jonsson_new_2013} and post HF equations
\cite{bostock_fully_2011}, convergence of which is strongly dependent
on the choice of the trial functions. Therefore, improving the choice
of the initial wave functions can significantly reduce the number of
iterations required for obtaining the desired solution \no{and,
  consequently, computation time} or enable the convergence at all.

At the same time a lot of effort has gone into improving the accuracy
of these models, while keeping their complexity low.  For example, the
relativistic corrections \cite{dolg_relativistic_2012,
  nakajima_douglaskrollhess_2012, marini_relativistic_1981,
  waber_relativistic_1975} and inhomogeneity
corrections~\cite{chau_systematic_2018} \no{were included into the TF
  model.}  However, solving the TF equation to produce numerically
stable results of electronic densities is in general nontrivial
problem and is an active area of research
\cite{akgul_constructing_2017, parand_accurate_2017,
  liu_laguerre_2015}.

Consequently, finding a set of analytical wave functions, which
provide better accuracy than the models described above would be
advantageous. As was recently demonstrated for nonrelativistic atoms
or ions, this can be achieved within the effective charge
model~\cite{skoromnik_analytic_2017}, where the hydrogen-like wave
functions can be used to analytically describe observable
characteristics with high accuracy. In this approach, one specifies a
complete and orthonormal hydrogen-like basis set with a single
parameter --- effective nuclear charge $Z^{*}$ --- identical for all
single-electron wave functions of a given atom. The basis completeness
allows one to perform a transition into the secondary quantized
representation, which is especially suited for the description of
many-body problems.

Then in order to compute observable characteristics one specifies the
set of occupation numbers for the state in question and the charge of
the nucleus. This allows one to determine the effective charge from
the vanishing first-order correction to the energy of the system
analytically. Then observable characteristics are expressed through
the expectation values, computed with the hydrogen-like wave functions
of this charge.

We point out here that perturbation theory in $Z^{-1}$ was developed
in many works (see e.g. \cite{vainshtein_energy_1985,
  fischer1977hartree} and citations therein), however, exactly the
introduction of the effective charge $Z^{*}$, instead of the usage of
the nuclear charge $Z$, significantly increases the accuracy of the
\no{leading-order} approximation, while keeping the complexity of all
calculations low as it contains the first-order correction implicitly.

As a result, it was demonstrated that the analytical \no{leading-order
  approximation describes} the whole spectrum of a nonrelativistic
multi-electron atom or ion and the associated perturbation theory
series was constructed, which converges fast with the rate of
$\sim 1/10$. The accuracy of these results does not depend on the
number of electrons in an atom, i.e., the effective charge \no{model}
is uniformly available for all states, both ground and excited ones,
of all atoms or ions. Moreover, the results via second-order
perturbation theory are comparable with those via multi-configuration
HF (MCHF) \cite{fischer1977hartree}.

We emphasize that the effective charge model is a completely
\textit{ab-initio} theory and thus crucially distinct from
semi-empirical models, such as those based on quantum defects of
Rydberg atoms \cite{foot2005atomic}, screened hydrogen
\cite{hau2012high} or empirical interpolation of X-ray measurements
\cite{genoni_can_2017}. In addition, contrary to numerical methods
such as solving the Hatree-Fock equations~\cite{hartree_1928}, the
effective charge model provides analytical expressions for electron
densities, scattering factors \no{and photoionization cross-sections.}

So far, the effective charge description has only been performed for
the nonrelativistic Schr\"{o}dinger equation. \change{However, it is
  well-known that leading-order relativistic corrections are
  proportional to~$(\alpha Z)^2$ and} \CHK{consequently}
\change{become important for heavy atoms and ions.}  Since the Dirac
equation for the hydrogen-like atom can be solved exactly and the
Dirac-Coulomb Green's function is known in closed form, it is possible
to develop a relativistic version of the effective charge
model. \CHK{As a result,} \no{this generalization is the main
  achievement of the present work.}

Similarly to the nonrelativistic case, we employ Dirac hydrogen basis
set with a single parameter --- the effective charge and require it to
be identical for all wave functions of a given atom. The basis
completeness allows us to work in the secondary-quantized
representation. We characterize states via electron occupation numbers
and determine the value of the effective charge, by requiring the
first-order perturbative correction to the energy of the system to
vanish. With this we obtain a description of energy spectra, both
ground and excited states together with wave functions. We have shown
that the analytical description within the effective charge model
agrees with the high accuracy with the numerical solution of
Dirac-Hartree-Fock (DHF) equations for all ions and atoms of the
periodic table.

The article is organized in the following way: in
Sec.~\ref{sec:relat-effect-charge} we introduce the effective charge
description and construct a perturbation theory series. In
Sec.~\ref{sec:zeroth-order-appr} we analyze the accuracy of the
zeroth- and first-order approximations. Consequently, we start from
computing numerical values of effective charges by solving
Eq.~(\ref{eq:13}). After this in Sec.~\ref{sec:ground-state-energ} we
determine the ground state energies for the first 100 atoms of the
periodic table and the electronic densities for some selected
atoms. In Sec.~\ref{sec:highly-charged-ions} we demonstrate that the
\no{leading-order} approximation is especially suited for the
description of highly charged ions, where our analytical functions are
almost coincident with numerical DHF wave functions. \no{In
  Sec.~\ref{sec:atom-scatt-fact} we analyze atomic scattering factors
  and in Sec.~\ref{sec:photo-ionis-cross} photoionization
  cross-sections. In Sec.~\ref{sec:conclusions}} we provide
conclusions and outlook. Finally, the details of the calculations are
summarized in Appendices
\ref{sec:appendix-a.-details}-\ref{sec:solut-relat-tf} and values of
the energies are given in Appendix \ref{sec:values-ground-state}.

\no{Atomic units with $\hbar = m = e
  = 1$ are employed throughout the paper.}

\section{Relativistic effective charge model}
\label{sec:relat-effect-charge}

\no{The Hamiltonian of the relativistic multi-electron atom can be
  written} in the secondary-quantized representation, using the
complete and orthonormal Dirac hydrogen-like basis set, with a single
parameter --- effective charge $Z^{*}$ --- identical for all wave
functions of the basis set. Consequently, the Hamiltonian of the
system reads
\begin{align}
      \hat{H}
  &= \no{\hat{H}_{0} + \hat{W}_{1} +
    \hat{W}_{2}},\label{eq:1}
  \\
  \hat{H}_0
    &=\int \hat{\Psi}^\dag(\vec{r}) \left(c\vec{\alpha} \cdot
      \hat{\vec{p}} + \beta c^{2} - \frac{Z^*}{|\vec{r}|}\right)
      \hat{\Psi}(\vec{r})d\vec{r}, \label{eq:2}
  \\
  \hat{W}_1
    &=\int  \hat{\Psi}^\dag(\vec{r}) \frac{Z^*-Z}{|\vec{r}|}
      \hat{\Psi}(\vec{r})d\vec{r}, \label{eq:3}
  \\
  \hat{W}_2
    &= \frac{1}{2}\int \frac{\hat{\Psi}^\dag(\vec{r})
      \hat{\Psi}^\dag(\vec{r'}) \hat{\Psi}(\vec{r})
      \hat{\Psi}(\vec{r'})}{|\vec{r}-\vec{r}'|}
      d\vec{r}d\vec{r'}, \label{eq:4}
\end{align}
where $c$ is the speed of light ($c \no{ = 1/\alpha}= 137.035999084$),
$Z$ the charge of the nucleus, $Z^{*}$ the effective charge, which
will be determined later, $\hat{\vec p} = -\ri \boldsymbol{\nabla}$
the momentum operator, $\vec \alpha$, $\beta$ the Dirac matrices and
$\hat\Psi(\vec r)$ the secondary-quantized operator~
\cite{FeynmanB1998statistical}. \no{We represent the perturbation
  operator as} a sum of single-electron $\hat{W}_1$ and
double-electron $\hat{W}_2$ operator components. \no{Moreover,} we do
not take into account the Breit part in the potential of the
electron-electron interaction and radiative quantum electrodynamics
corrections, thus limiting ourselves to the description of the
multi-electron atom within DHF approximation. However, we mention here
that for large $Z$ these corrections become relevant and would have to
be included for all models. We stress here that Eq.~(\ref{eq:1}) is
the exact transformation of the Hamiltonian of the multi-electron
atom, since we just added and subtracted the term proportional to
$Z^{*}$. \no{The eigenvalues of $\hat{H}$ define the values of the
  energy $E$ of a multi-electron atom.}

The secondary-quantized \no{operator} $\hat{\Psi}(\vec r)$ is expanded
in the complete Dirac hydrogen-like basis, which is dependent on
$Z^{*}$
\begin{equation}
  \hat{\Psi}(\vec r) = \sum_\nu\left( \hat{a}_\nu
    \Psi_\nu^\epsilon(\vec r\no{,Z^*}) + \hat{b}^\dag_\nu
    \Psi_\nu^{-\epsilon}(\vec r\no{,Z^*})\right), \label{eq:5}
\end{equation}
with $\Psi^\epsilon_{\nu}(\vec r\no{,Z^*})$ and
$\Psi^{-\epsilon}_{\nu}(\vec r\no{,Z^*})$ being the positive and
negative energy \no{eigenfunctions} of the single-particle Dirac
Hamiltonian respectively, identified by the \no{set of} collective
quantum numbers $\nu=n_r l j m_{j}$ and $\nu = \vec p l j m_{j}$ for
the discrete and continuous spectra correspondingly (actual
expressions for the functions are given in Appendix
\ref{sec:appendix-a.-details}). The fermionic operators of positive
\no{($\hat{a}_\nu$) and of negative ($\hat{b}_\nu$)} energy satisfy
the standard anticommutation relations
\no{$\{\hat{a}_\nu, \hat{a}^\dag_{\nu'}\}=\{\hat{b}_\nu,
  \hat{b}^\dag_{\nu'}\}=\delta_{\nu,\nu'}$}, which ensures that
multi-particle states $|\nu_1...\nu_N\rangle$ correspond to fully
antisymmetric Slater determinants. In the following, we will denote
electron occupation numbers with $\lambda_{i}$ and \no{negative
  energy} occupation numbers as $\mu_{i}$.

In order to compute the energy of the system we specify a set of $N$
occupation numbers $\lambda_{1}, \ldots, \lambda_{N}$, which
characterizes the state of $N$ electrons
$|\lambda_{1},\ldots,\lambda_{N}\rangle$ and construct a perturbation
series by considering the operator
$\hat{W} = \hat{W}_{1} + \hat{W}_{2}$ as a perturbation
\begin{align}
  E
  &= E^{(0)} (Z^*) + \Delta E^{(1)} (Z^*) + ... , \label{eq:6}
  \\
  \Delta E^{(1)} (Z^*)
  &= \langle\lambda_{1},\ldots,\lambda_{N}| \hat{W}
    |\lambda_{1},\ldots,\lambda_{N}\rangle. \label{eq:7}
\end{align}

Since the state $|\lambda_{1},\ldots,\lambda_{N}\rangle$ is the
eigenstate of $\hat{H}_{0}$, the zeroth-order energy $E^{(0)}(Z^{*})$
is a sum of hydrogen-like Dirac energies over the occupied states. The
expression for the Dirac energy of the hydrogen-like atom is
well-known \cite{flugge_practical_1971} and is given by
\begin{equation}
  E^D_{n_{r} j} (Z^*) = \frac{c^{2}}{\sqrt{1 +
      \left(\frac{\alpha Z^{*}}{n_{r} + \sqrt{(j+1/2)^{2} - (\alpha
            Z^{*})^{2}}}\right)^{2}}}. \label{eq:8}
\end{equation}
For this reason the energy $E^{(0)}(Z^{*})$ reads
\begin{align}
  E^{(0)}(Z^{*}) = \sum_{{\lambda_{i}}}
  E_{n_{r}jm_{j}}^{D}(Z^{*}). \label{eq:9} 
\end{align}

Evaluation of the first-order correction to the energy of the
system is straightforward:
\begin{align}
  \Delta E^{(1)} (Z^*)
  &= \langle \lambda_{1},...,\lambda_{N} | \hat{W}
    |\lambda_{1},...,\lambda_{N}\rangle \nonumber
  \\
  &= (Z^*-Z)\sum_{k = 1}^{N} A_{\lambda_k}(Z^*) \label{eq:10}
  \\
  &+ \sum_{k<l = 1}^{N}
    \left(B_{\lambda_l,\lambda_l}^{\lambda_k,\lambda_k}(Z^*) -
    B_{\lambda_k,\lambda_l}^{\lambda_l,\lambda_k}(Z^*)\right),
    \nonumber
\end{align}
where we have defined:
\begin{align}
  A_{\lambda_{k}}
  &= \int
    \frac{|\Psi_{\lambda_{k}}^{\epsilon}(\vec{r})|^2}{|\vec{r}|}
    d\vec{r}, \label{eq:11}
  \\
  B_{\lambda_2,\lambda_4}^{\lambda_1,\lambda_3}
  &=
    \int\frac{\Psi_{\lambda_1}^{\epsilon*}(\vec{r})
    \Psi^{\epsilon*}_{\lambda_2}(\vec{r}')
    \Psi_{\lambda_3}^{\epsilon}(\vec{r})
    \Psi_{\lambda_4}^{\epsilon}(\vec{r}')}{|\vec{r}-\vec{r}'|}
    d\vec{r} d\vec{r}', \label{eq:12}
\end{align}
with the implied dependence on the effective charge omitted for
simplicity. The $A_{\lambda_{k}}$ describes the contribution from the
single-particle operator $\hat{W}_{1}$, while
$B_{\lambda_2,\lambda_4}^{\lambda_1,\lambda_3}$ from the
double-particle one $\hat{W}_{2}$. The last term in Eq.~(\ref{eq:10})
is the difference of the Coulomb and exchange integrals. Both
$A_{\lambda_{k}}$ and $B_{\lambda_2,\lambda_4}^{\lambda_1,\lambda_3}$
can be evaluated analytically for a given set of occupation numbers
(See Appendix \ref{sec:appendix-a.-details}).

\begin{figure}
  \includegraphics[width=80mm]{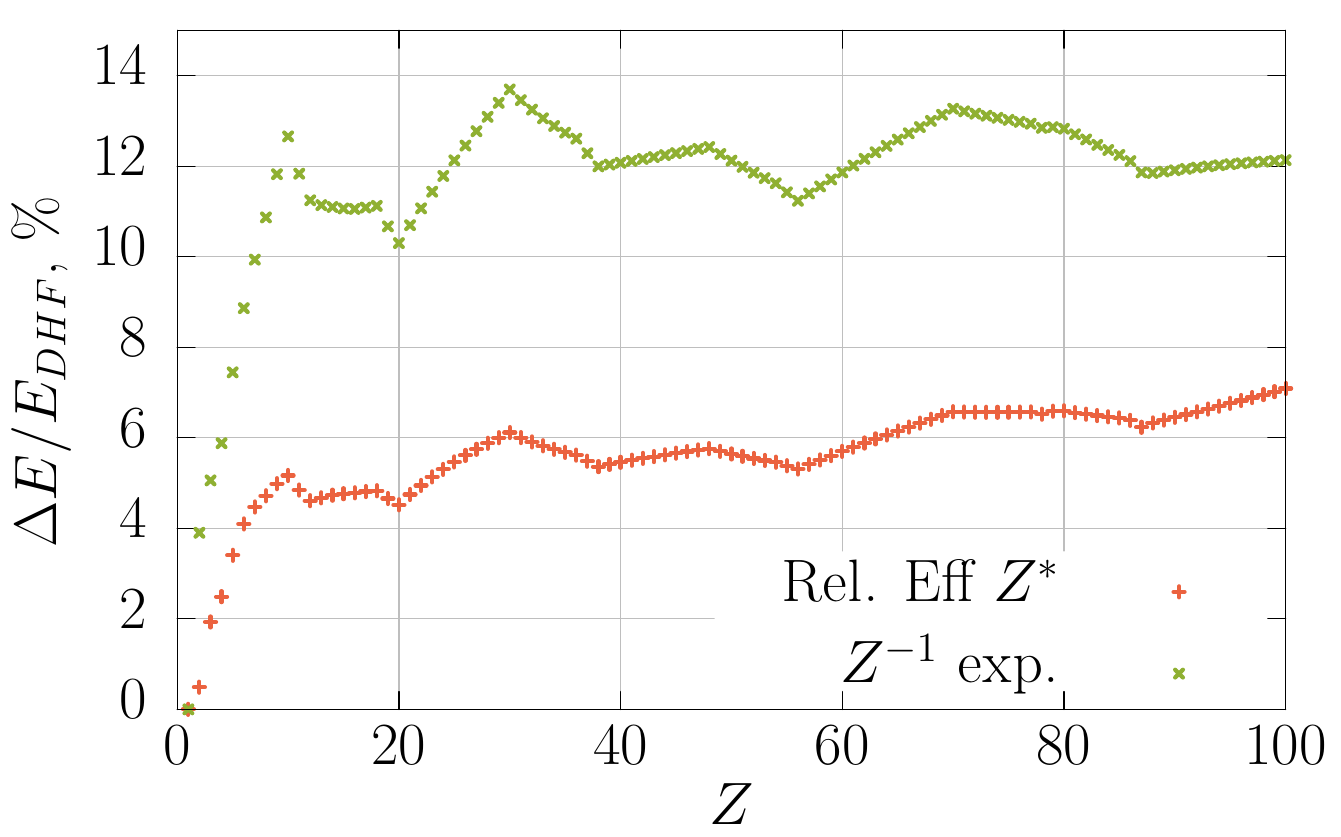}
  \caption{(color online) The relative difference between the values
    of ground state energies of the first 100 neutral atoms obtained
    via the effective charge model ($E^{(0)}$) and $1/Z$ expansion to
    Dirac-Hartree-Fock values
    ($E_{\mathrm{DHF}}$)\cite{DESCLAUX1973311}. \no{See
      Sec. \ref{sec:ground-state-energ} for details.}}
  \label{accuracy}
\end{figure}

In addition, we mention here that fermionic operators of negative
energies do not contribute to the energy of the system in the zeroth-
and first-order as we are considering the corrections only to the
electronic states, that is the states that do not contain $\mu_{i}$
occupation numbers. However, starting from second-order perturbation
theory, the negative energy states will contribute to the observable
characteristics, since there exist nonvanishing matrix elements due to
the structure of the interaction operator $\hat{W}$.
\begin{figure*}
  \centering  
  a)\subfigure{\includegraphics[height=43mm]{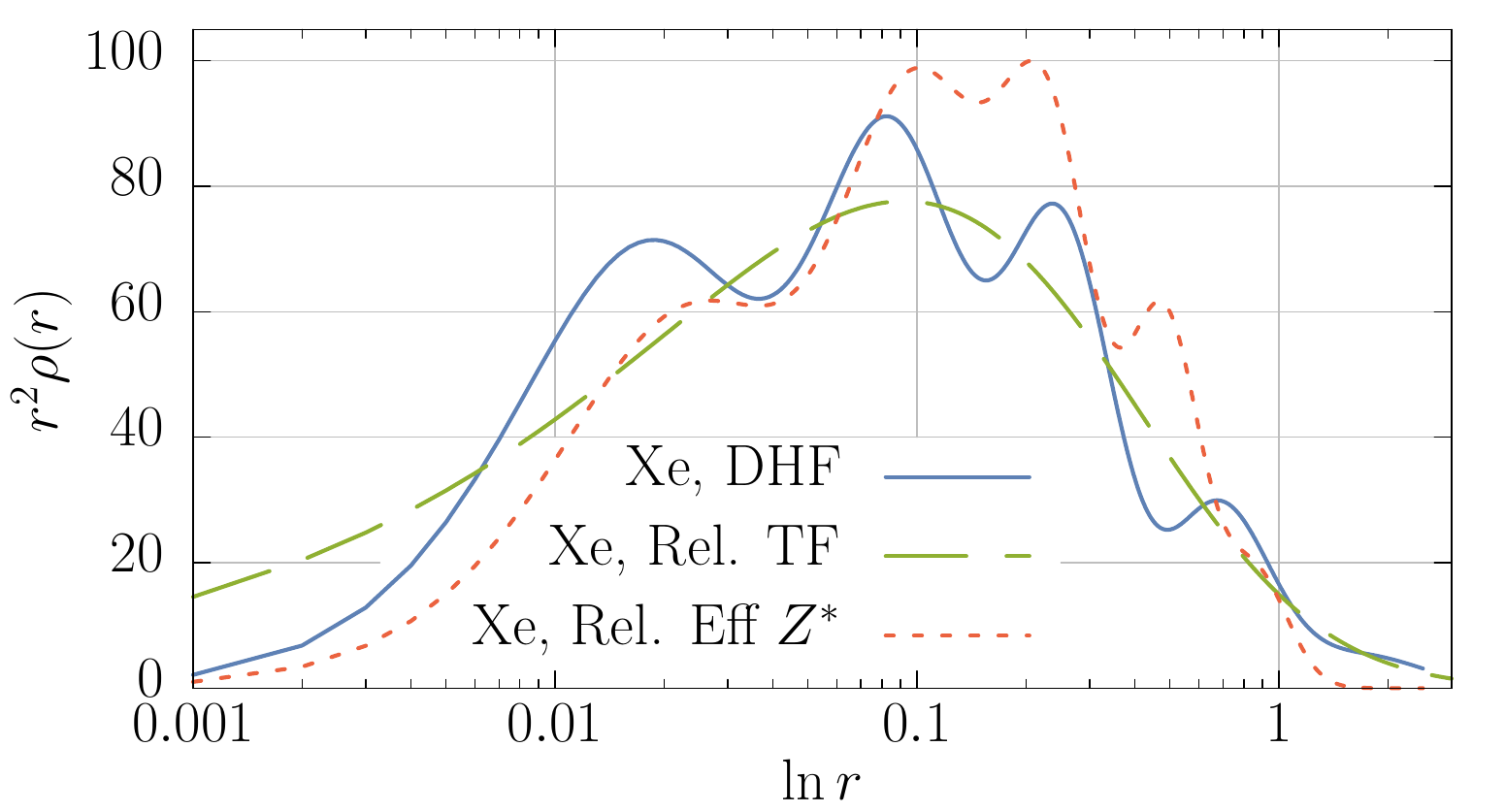}}
  b)\subfigure{\includegraphics[height=43mm]{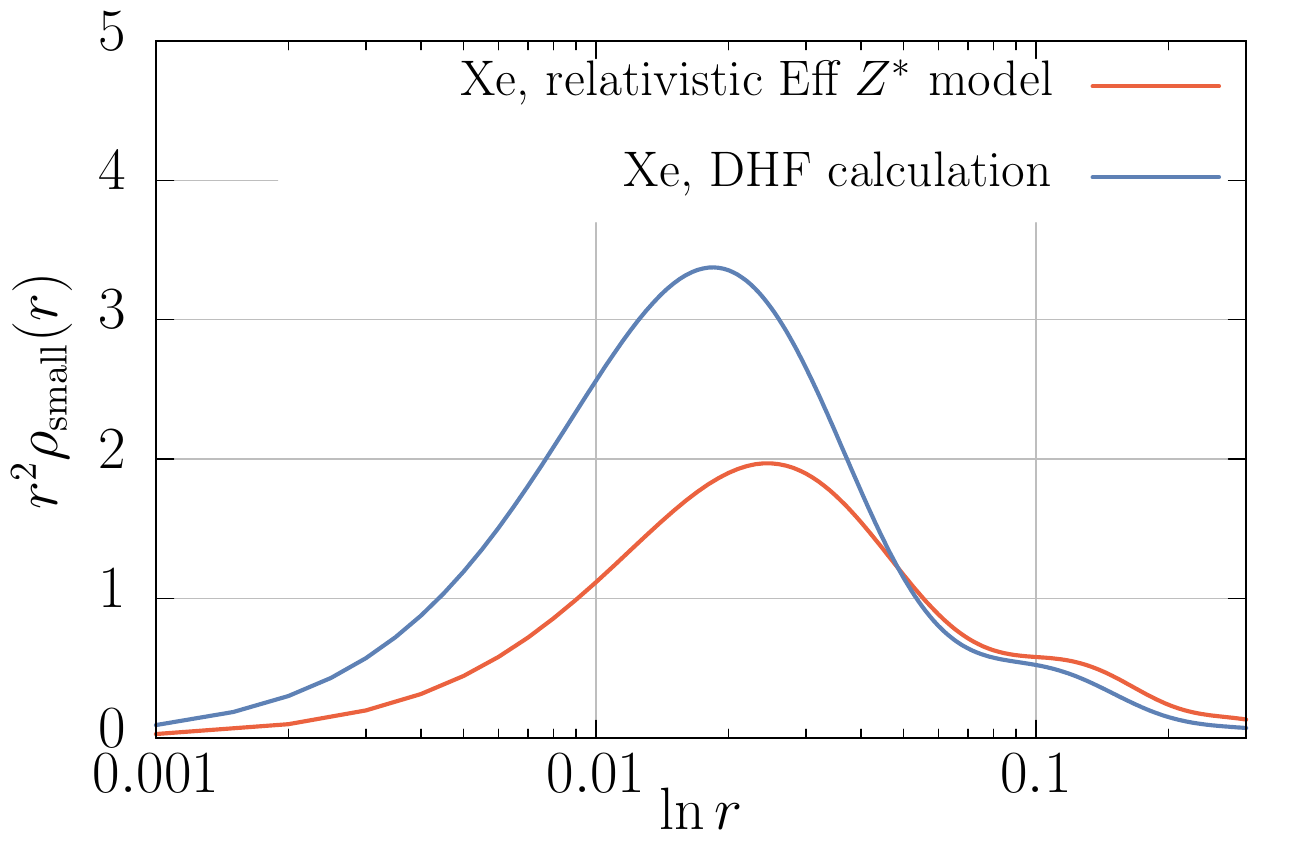}}
  c)\subfigure{\includegraphics[height=43mm]{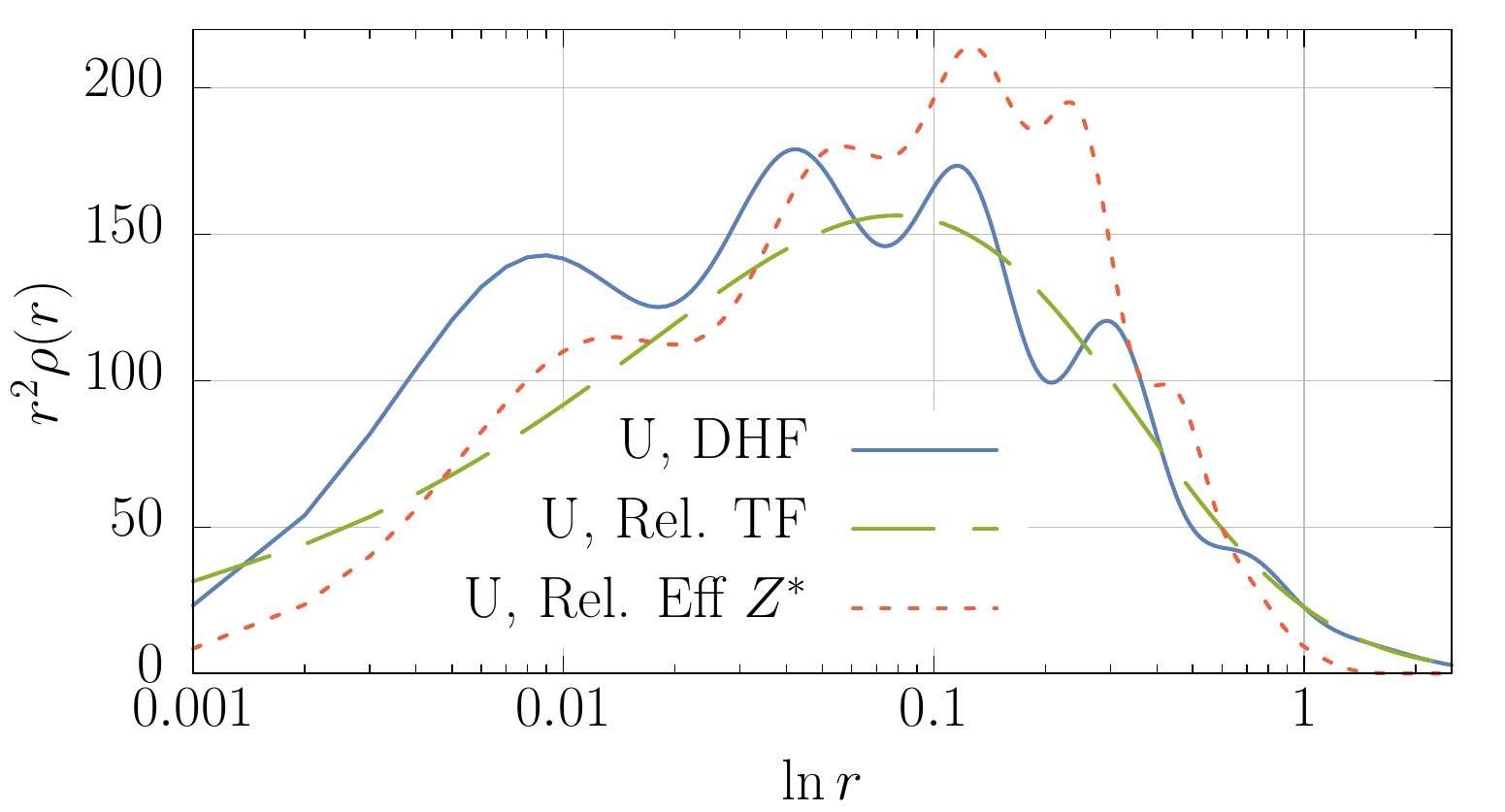}}
  d)\subfigure{\includegraphics[height=43mm]{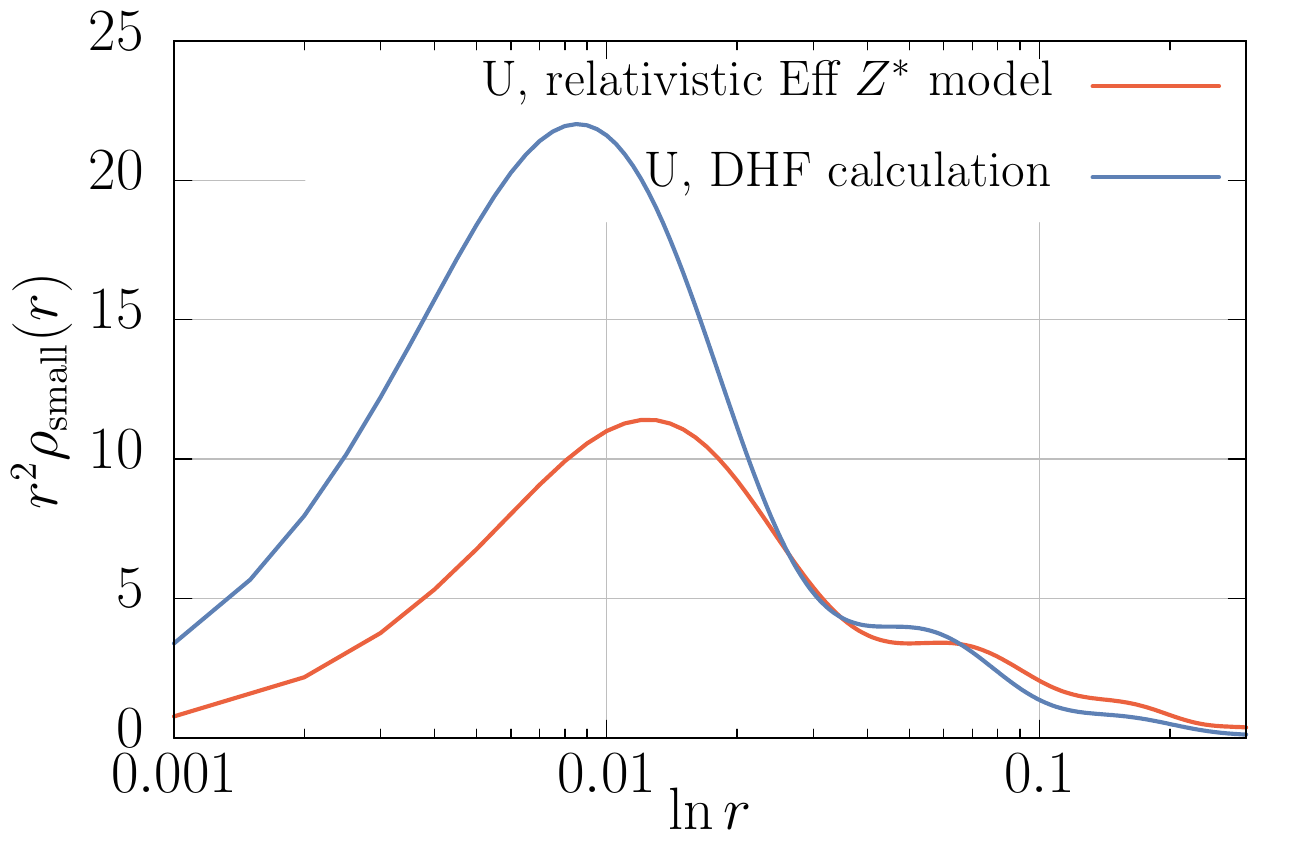}}
  \caption{(color online)\no{(a and c)} Total radial electronic
    densities and \no{(b and d)} densities \no{composed from the
      small} components of the \no{Dirac wave functions
      ($\sum_{k=1}^{N}|F_{\lambda_{k}}(r,Z^{*})|^{2}$, where the
      definition of $F_{\lambda_{k}}(r, Z^{*})$ is given in appendix
      \ref{sec:appendix-a.-details}).} Figures a and b represent
    neutral xenon \no{Xe${}^{54}$}, c and d neutral uranium
    \no{U${}^{92}$}. Dashed, red line is the zeroth-order effective
    charge approximation, while blue and green, long-dashed lines are
    the results of the numerical solution of DHF (obtained via GRASP2k
    \cite{jonsson_new_2013, DYALL1989425}) and relativistic TF (see
    Appendix~\ref{sec:solut-relat-tf}) equations,
    respectively. \no{See Sec. \ref{sec:ground-state-energ} for
      details.}}
  \label{fig:densities_neutral_atoms}
\end{figure*}

To find the effective charge we proceed in analogy with
Ref.~\cite{skoromnik_analytic_2017} and choose it from the condition
that the first-order correction to the energy of the system for a
given state is vanishing, i.e.,
\begin{align}
  \Delta E^{(1)}(Z^*) = 0. \label{eq:13}
\end{align}
For this reason the expression for the energy of the system in
first-order perturbation theory is given via a sum of hydrogen-like
energies, Eq.~(\ref{eq:9}) with the effective charge $Z^{*}$, defined
as a solution of Eq.~(\ref{eq:13}).

It is worth noting here, that the nontrivial dependence of Dirac
hydrogen wave functions on the nuclear charge makes it impossible to
separate the effective charge from the above integrals. \no{This means
  that} contrary to the nonrelativistic case, $A_{\lambda_{k}}$ and
$B^{\lambda_{1},\lambda_{3}}_{\lambda_{2},\lambda_{4}}$ are implicitly
dependent on $Z^*$. This is related to the fact that the Dirac
equation, unlike the Schr\"{o}dinger equation, is not scale
invariant. In fact, rescaling the radial variable $r$ in a
Schr\"{o}dinger hydrogen atom, effectively changes its charge, while
in the Dirac hydrogen atom, it effectively changes its mass.

\change{
  For this reason Eq. (13) can only be solved approximately. In
  practice it means finding a root of an equation containing gamma
  functions. For example, the relativistic effective charge $Z^*$ of a
  Helium-like atom or ion with nuclear charge $Z$ is found by solving:
  \begin{equation}
    2(Z^*-Z)+1 = \frac{\Gamma(2\gamma+1/2)}{\Gamma(2\gamma+1)\sqrt{\pi}},
  \end{equation}
  where $\gamma = \sqrt{1-(\alpha Z^*)^2}$. Such equation can be
  solved to any order of $\alpha$ with traditional iterative methods
  from numerical analysis or with analytical approximations. In the
  latter case, the Taylor series of the Gamma function can be used to
  approximate the effective charge to any order in $\alpha$. Up to the
  second order in $\alpha$ it reads:
  \begin{equation*}
    Z^* = Z-\frac{5}{16}+\alpha^2 \left(Z-\frac{5}{16}\right)^2
    \frac{12 \log(2)-7}{32}+O[\alpha^4].
  \end{equation*}
}

Following the procedure of calculating the effective \no{charge of the
  nucleus}, we can easily obtain expressions for electronic densities
of any atom or ion. Using the density operator:
$\hat{\rho}(\vec r) = \hat{\Psi}^\dag(\vec r\no{,Z^*})\hat{\Psi}(\vec
r\no{,Z^*})$, we get the zeroth-order density of a given
multi-electron state as:
\begin{align}
  \rho^{(0)}(\vec r)
  &= \langle \lambda_1, \ldots, \lambda_N | \hat{\rho}(\vec{r})
    |\lambda_1, \ldots, \lambda_N\rangle \nonumber
  \\
  &= \sum_{k=1}^{N}
    |\Psi_{\lambda_k}^{\epsilon}(\vec{r}\no{,Z^*})|^2, \label{eq:14}
\end{align}
with the explicit expressions for hydrogen-like wave functions
$\Psi_{\lambda_k}^{\epsilon}(\vec r\no{,Z^*})$ given in Appendix
\ref{sec:appendix-a.-details}. In fact, \no{Eq.~}(\ref{eq:14})
represents very simple products of exponentials and polynomials, thus
allowing its use in numerical plasma \cite{chung_flychk:_2005} or
description of ionization in particle-in-cell (PIC) computer codes
\no{for} laser-matter interactions \cite{arber_contemporary_2015}.

\begin{figure*}
  \centering  
  \subfigure{\includegraphics[width=80mm]{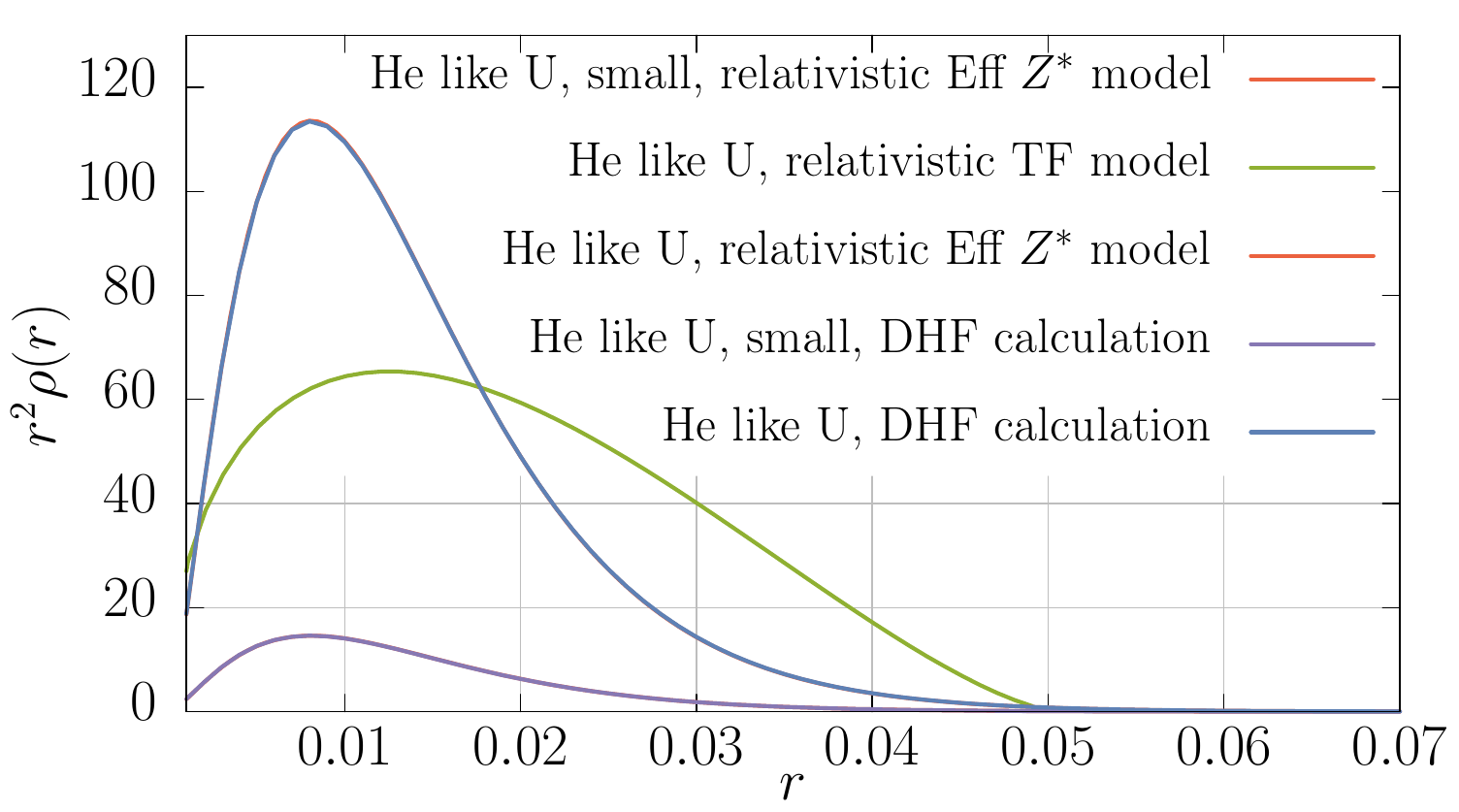}}
  \subfigure{\includegraphics[width=80mm]{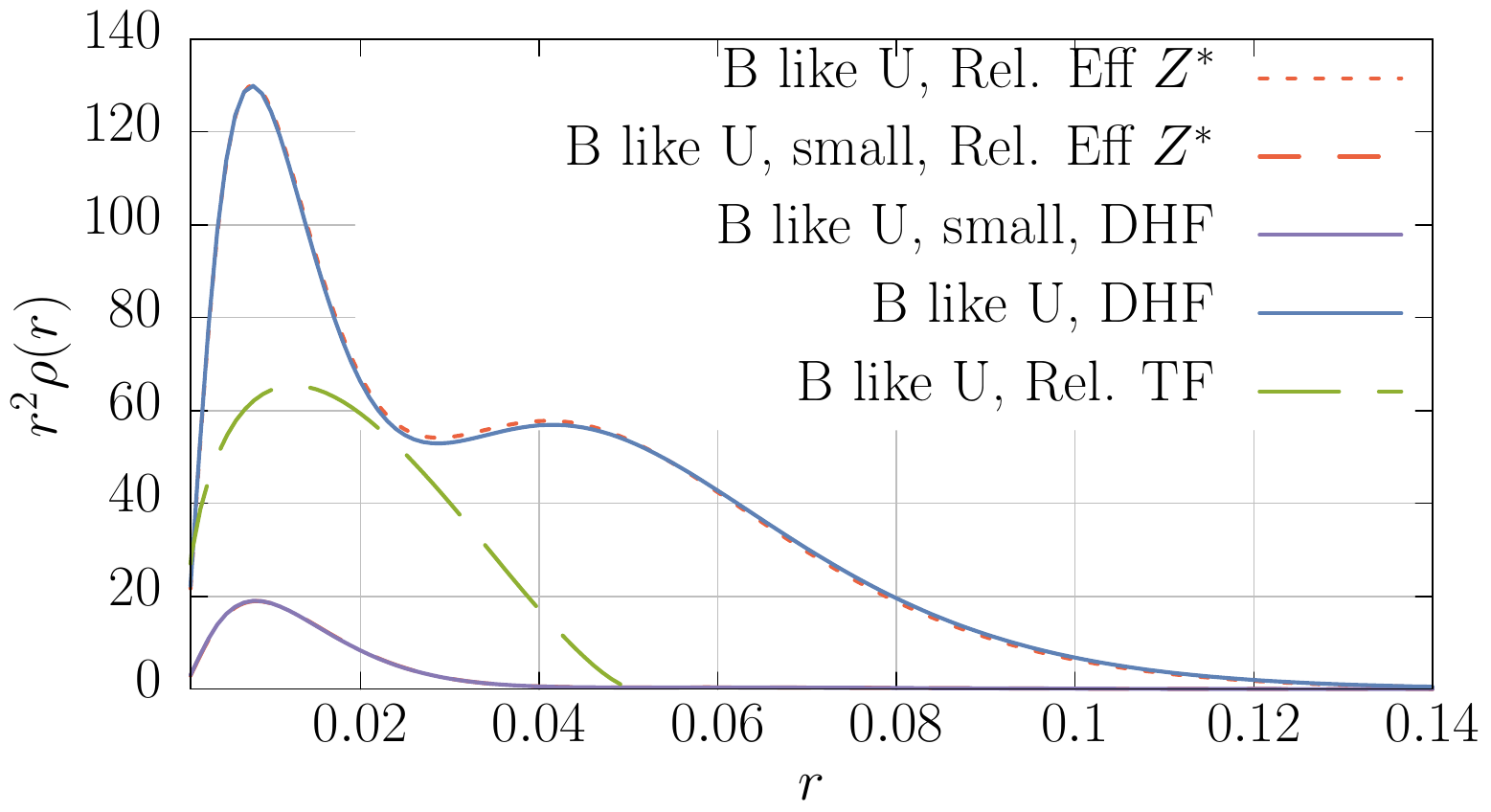}}
  \caption{(color online) Total and \no{small} component electronic
    densities of highly charged uranium \no{U${}^{92}$}. Dashed, red
    line is the zeroth-order effective charge model (completely
    overlaps the DHF result), while blue solid and green long-dashed
    lines are results of the DHF (obtained via GRASP2k
    \cite{jonsson_new_2013, DYALL1989425}) and TF models respectively
    (see Appendix~\ref{sec:solut-relat-tf}). \no{See
      Sec. \ref{sec:highly-charged-ions} for details.}}
  \label{fig:HCI}
\end{figure*}

\section{Accuracy of the zeroth-order approximation}
\label{sec:zeroth-order-appr}

\subsection{Ground state energies and electronic
  densities of neutral atoms}
\label{sec:ground-state-energ}

In order to calculate the ground-state energies of multi-electron
atoms or ions, one needs to specify a set of occupation numbers,
characterizing the state within our Dirac hydrogen-like
basis. However, in the zeroth-order approximation the particular
choice may not have the correct ordering due to relatively low
ionization energies of heavy atoms.

The comparison of our analytical expressions obtained via
Eq.~(\ref{eq:9}) for the values of ground state energies with the
results obtained via solutions of DHF equations demonstrates that
similarly to the nonrelativistic case, the optimal choice of the
occupation numbers is given according to the ``Aufbau'' or
Madelung–Janet–Klechkovskii rule \cite{Madelung1936,
  KlechkovskiiA1962justification, doi:10.1021/ed056p714}, as it
provides a simple and consistent choice of occupation numbers for any
number of electrons, while resulting in the lowest zeroth-order ground
state energies in almost all cases. For example, from the three sets
of occupation numbers $[\mathrm{Xe}]6\mathrm{s}^{1}$,
$[\mathrm{Xe}]4\mathrm{f}^{1}$ and $[\mathrm{Xe}]5\mathrm{d}^{1}$ for
\no{cesium Cs${}^{55}$}, \no{according to the results of our
  relativistic effective charge model,} the first state possesses the
lowest energy, while the remaining two are excited states. Compare the
values for the energy
$E_{[\mathrm{Xe}]6\mathrm{s}^{1}} = -7361.18$~a.u versus
$E_{[\mathrm{Xe}]4\mathrm{f}^{1}} = -7346.19$~a.u and
$E_{[\mathrm{Xe}]5\mathrm{d}^{1}} = -7354.40$~a.u, which corresponds
to the ``Aufbau'' rule \cite{Madelung1936,
  KlechkovskiiA1962justification, doi:10.1021/ed056p714}.

\change{This shows that despite using a single value of effective
  charge for all electrons, the errors coming from internal and outer
  electrons compensate each other, providing} \CHK{sufficient}
\change{accuracy to describe the correct order} \CHK{for the}
\change{filling of atomic shells. This is a principal difference of
  the proposed model from other single-parametric models such as}
\CHK{the} \change{Thomas-Fermi or $Z^{-1}$ expansion.}

The accuracy of the zeroth-order calculation is presented in
Fig.~\ref{accuracy} and numerical values of effective charges and
energies are given in Appendix~\ref{sec:values-ground-state} in
Tab.~\ref{tab:ground_state_energies}. We compare our results with
solutions of DHF equations from Ref.~\cite{DESCLAUX1973311}. As can be
concluded from Fig.~\ref{accuracy} the effective charge description
leads to a uniform approximation, i.e., the relative accuracy is
independent of the number of electrons of an atom and $\sim 6\%$ with
respect to DHF for all elements of the periodic table. This is
considerably better than the TF approximation
\cite{landau1981quantum}. \change{Furthermore, based on the results of
  the non-relativistic effective charge model calculations
  \cite{skoromnik_analytic_2017}, we expect that the accuracy can be
  improved by at least one order of magnitude by the inclusion of
  second-order corrections.}

We point out here that if we take the energy of the system in
first-order perturbation theory and consider the effective charge to
be equal to the full nuclear charge, one obtains the perturbation
theory over $Z^{-1}$. \change{In Fig.~\ref{accuracy} we present a
  comparison of the values of total binding energies} from the
effective charge model with the ones from $Z^{-1}$
expansion. \change{It is apparent that the effective charge model
  gives much more accurate results than the $Z^{-1}$ expansion.} For
example, for He one gets $-2.75$ a.u vs $-2.86$ a.u using $Z^{-1}$
expansion and effective charge model respectively, while for larger
atoms the accuracy drops significantly.

\begin{figure*}
  \centering  
  \subfigure{\includegraphics[width=80mm]{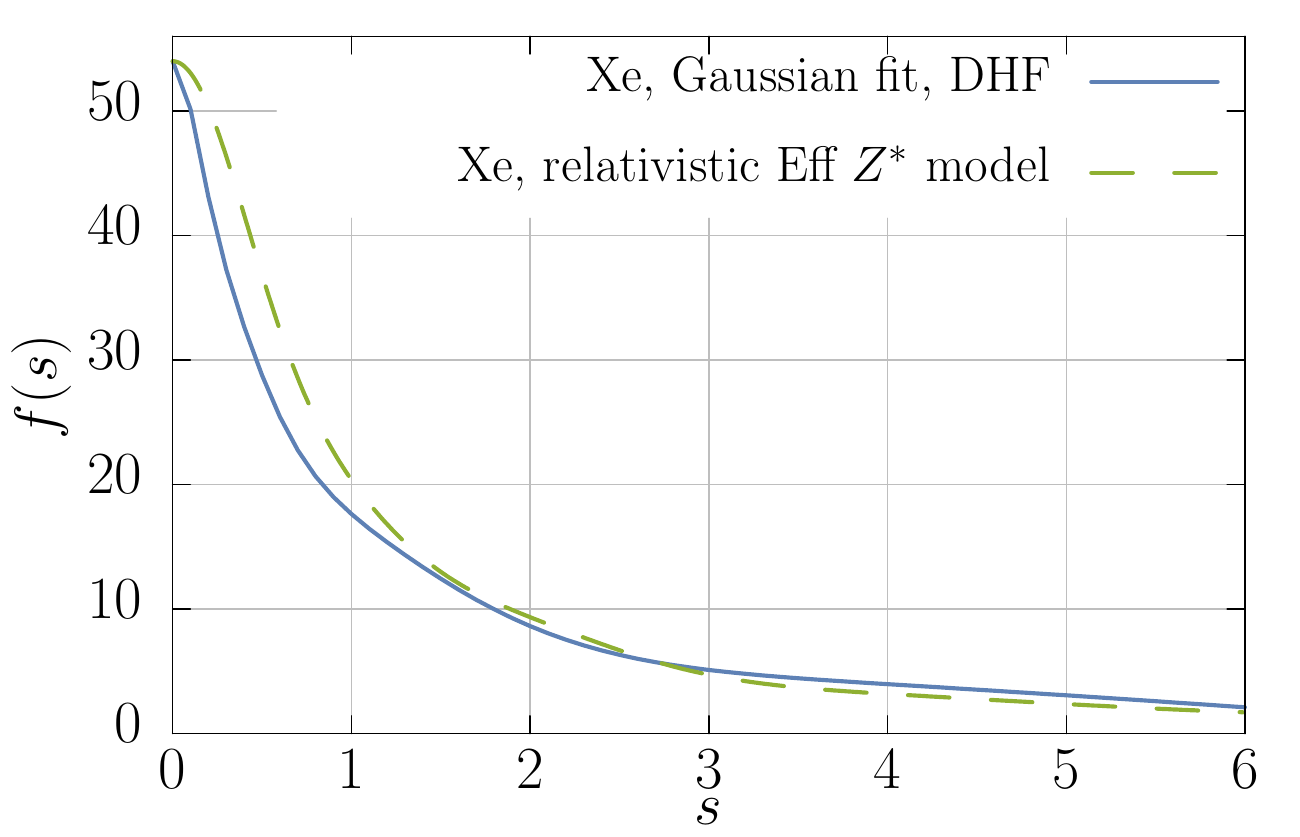}}
  \subfigure{\includegraphics[width=80mm]{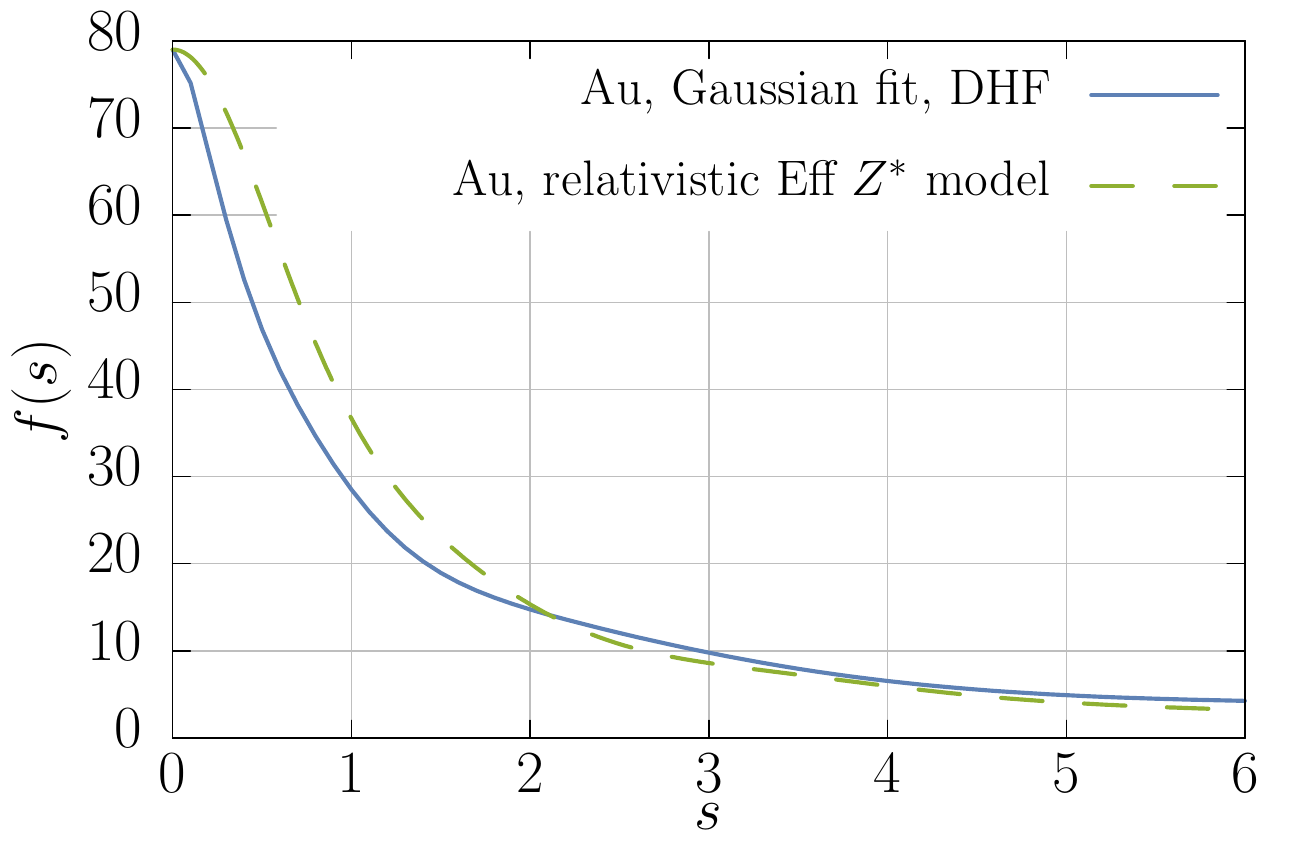}}
  \subfigure{\includegraphics[width=80mm]{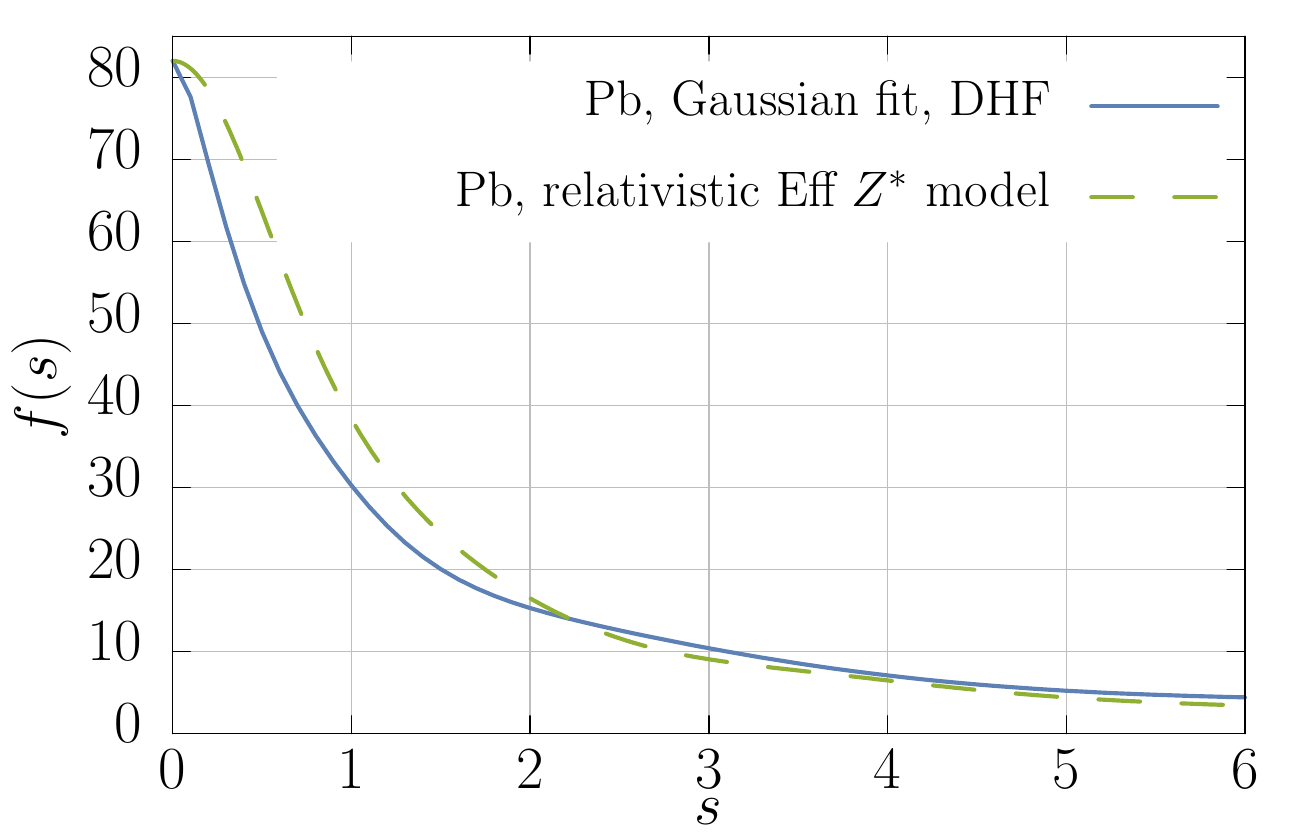}}
  \subfigure{\includegraphics[width=80mm]{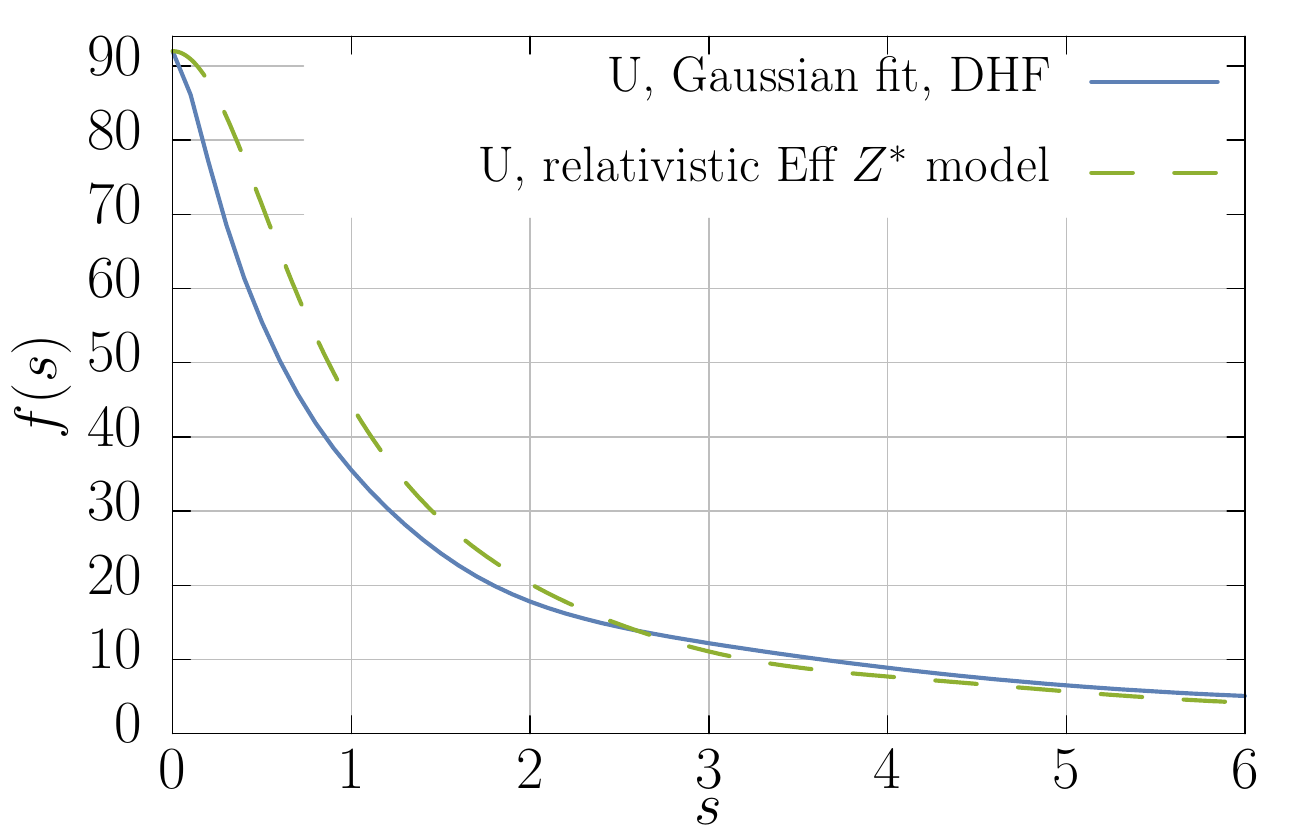}}
  \caption{(color online) Atomic scattering factors of neutral
    \no{xenon Xe$^{54}$, gold Au$^{79}$, lead Pb$^{82}$ and uranium
      U$^{92}$ atoms}, as a function of the scattering parameter
    $s = \sin\theta / \lambda,\ [\textup{\AA}^{-1}]$, with $\theta$
    being Bragg's angle. The quantity $s$ is related to the absolute
    value of $q$ from Eq.~(\ref{eq:15}) as
    $q = 4\pi s \cdot 0.529177$. Dashed, green line is the analytic
    result via Eq.~(\ref{eq:15}) of the effective charge model, while
    the solid blue curve is a Gaussian fit of DHF from
    Ref.~\cite{waasmaier_new_1995}. \no{See
      Sec. \ref{sec:atom-scatt-fact} for the detail.}}
  \label{fig:ASF}
\end{figure*}

In order to compute the electronic density, one needs to use
Eq.~(\ref{eq:14}), i.e., to compute the sum of squares of absolute
values of Dirac hydrogen-like wave functions. After simplification,
this sum is just a product of exponentials and
polynomials. Consequently, our \no{model}, unlike the DHF and TF
calculations, provides fully analytical expressions for electronic
densities and is therefore particularly useful for applications
requiring repeated calculations. Relatively simple expressions
resulting from our model can, therefore, be incorporated into existing
software, used in the description of X-ray scattering on M\"{o}ssbauer
crystals \cite{Sturhahn2000}.

In Fig.~\ref{fig:densities_neutral_atoms} we plot the resulting
dependence of electronic densities on the radial coordinate $r$ for
selected neutral atoms. \change{Despite the fact that the effective
  charge model underestimates the density for high $r$, it agrees well
  with} \CHK{the} \change{DHF result already in the zeroth-order
  approximation. Contrary to the relativistic TFD model, it correctly
  reproduces all of the qualitative features, including all density
  oscillations and} \CHK{the} \change{overall asymptotic
  behaviour}. In addition, we point out that the relativistic TFD
model \cite{dolg_relativistic_2012, nakajima_douglaskrollhess_2012,
  marini_relativistic_1981, waber_relativistic_1975}, unlike its
nonrelativistic counterpart \cite{landau1981quantum}, does not have a
universal dependence on the charge of the nucleus. For this reason, in
order to obtain the electronic density, the relativistic TFD equation
needs to be repeatedly solved numerically, which is a nontrivial
procedure. (See Appendix \ref{sec:solut-relat-tf}).

\subsection{Highly charged ions}
\label{sec:highly-charged-ions}

The effective charge model is \no{also suitable for the description of
  ions,} since it does not require the \no{charge of the nucleus} to
be equal to the number of electrons of an atom. Consequently, we fix
the nuclear charge as $Z$ and use a state vector
$|\lambda_{1},\ldots,\lambda_{N}\rangle$, where $N \neq Z$. As an
example we evaluated energies of ground and excited states of He-, Li-
and B-like uranium ions \no{U$^{90+}$, U$^{89+}$ and U$^{87+}$,
  respectively}.

We \no{would like} to stress here again, that the effective charge
$Z^{*}$ is identical for all single particle states from a set that
specifies the state. However, two different states will have different
charges. For example, the He-like uranium \no{U$^{90+}$} ground state
$1s_{\uparrow}1s_{\downarrow}$ and the excited state
$1s_{\uparrow}2s_{\downarrow}$ possess \no{unequal} effective charges.

The results of the calculation of electronic densities are presented
in Fig.~\ref{fig:HCI}. The electronic densities, despite being
analytical expressions in the zeroth-order approximation, coincide
remarkably well with the ones obtained from numerical solutions of the
DHF equations.

Furthermore, the energy eigenvalues are presented in
Tab.~\ref{tab:Ions}. It is clear that the accuracy is significantly
better (below $0.02\%$) than for neutral atoms and in fact sufficient
to obtain correct ordering, even for very closely spaced excited
states. It is worth pointing out that the accuracy can be further
improved by forming linear combinations of all sub-configurations
arising from the single $JLS$ configuration and subsequently
diagonalizing the Hamiltonian exactly in such finite basis --- a
technique that has been successfully employed in the nonrelativistic
case \cite{skoromnik_analytic_2017}.  \no{We also compared our
  effective-charge model results with the results of
  configuration-interaction Dirac-Fock-Sturm method \cite{Tup2003OS},
  presented in the last column of Table~I.}

Finally, we note that $Z^{-1}$ expansion gives reasonable results for
highly charged ions. However, it is less accurate than the effective
charge model by at least $50\%$.

\begin{figure*}
  \centering  
  \subfigure{\includegraphics[width=80mm]{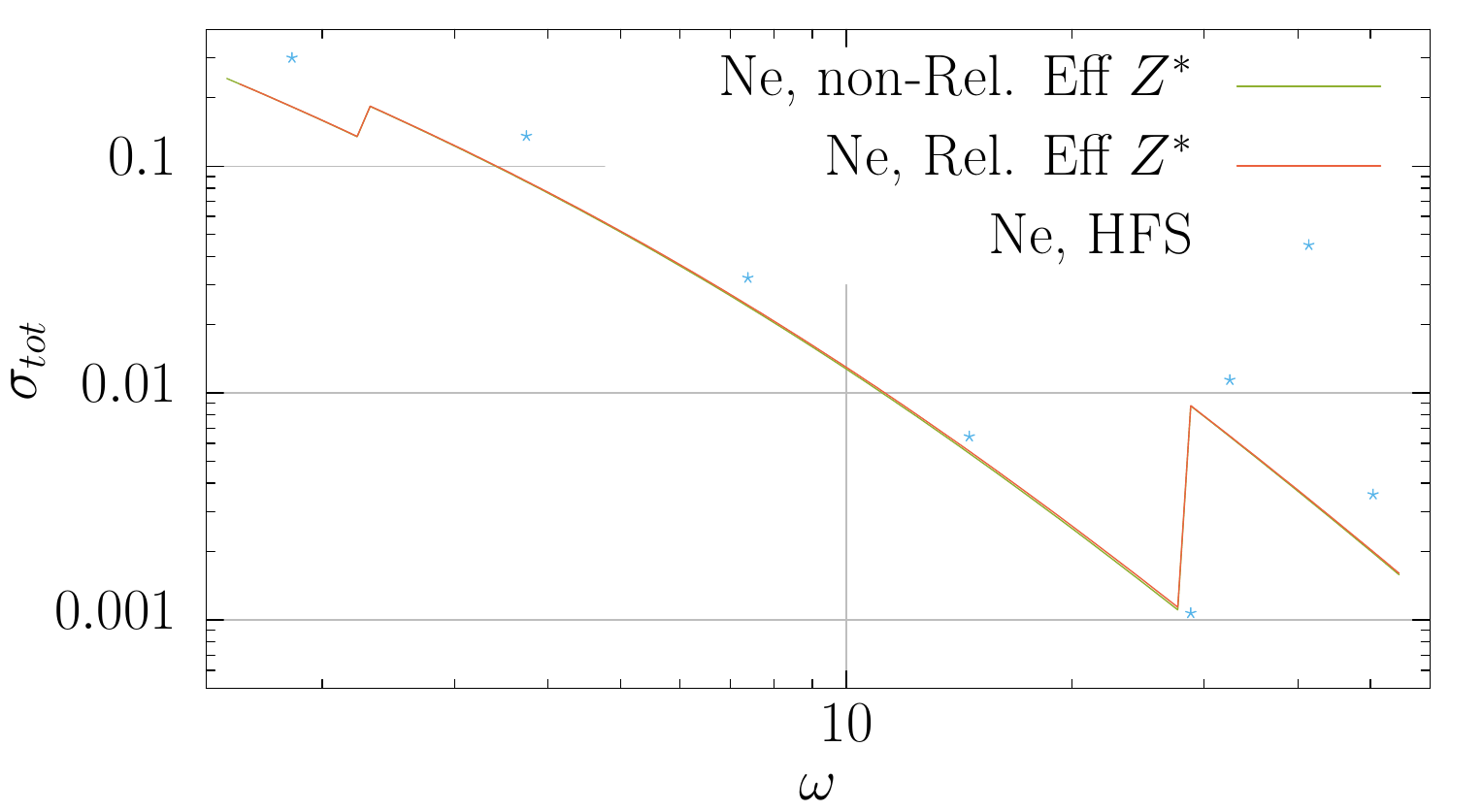}}
  \subfigure{\includegraphics[width=80mm]{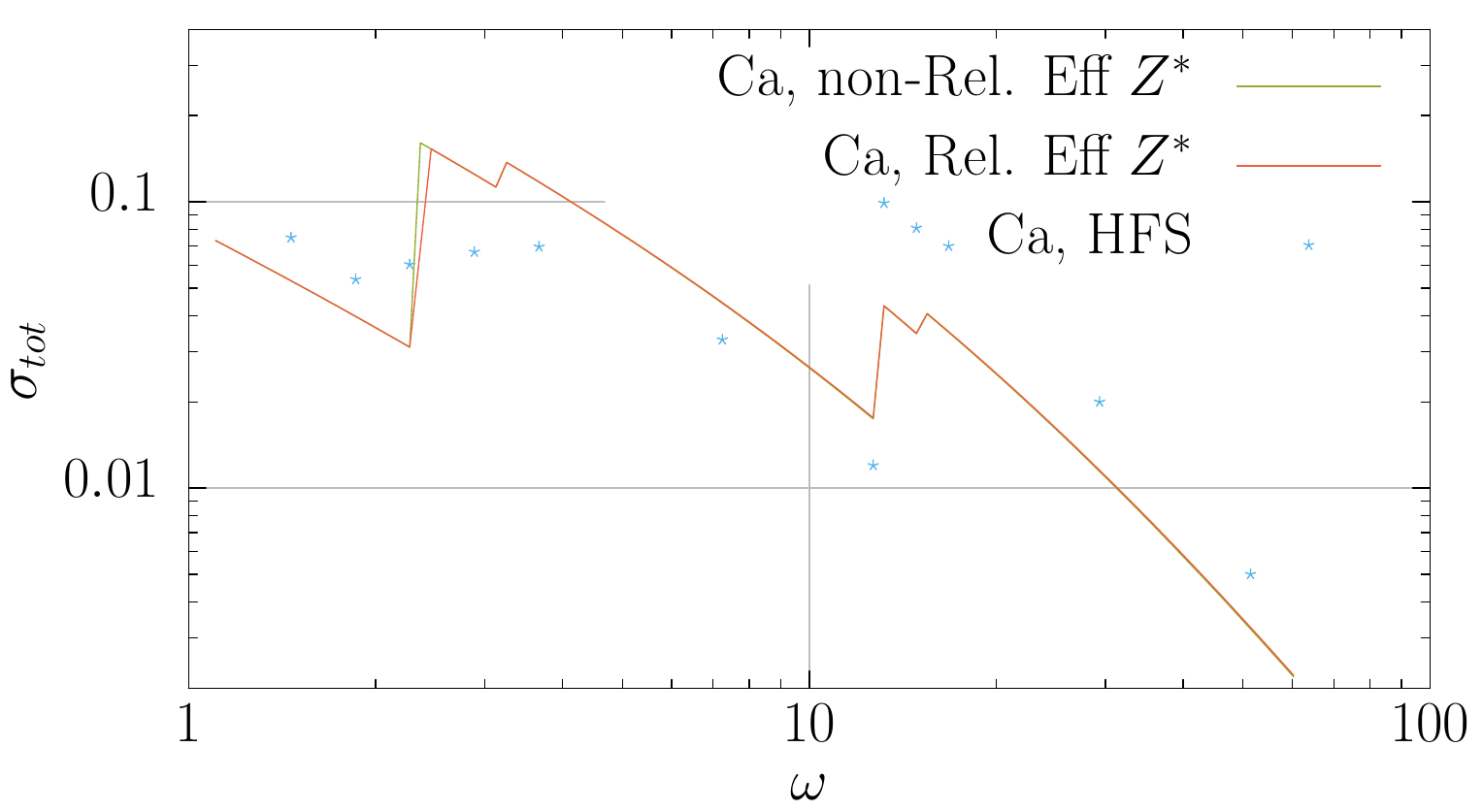}}
  \caption{\change{(color online) Comparison of the non-relativistic
      (green) and relativistic (red) effective charge model
      calculation of the total photoionization cross-section for neon
      Ne$^{10}$ (left) and calcium Ca$^{20}$
      (right). Hartree-Fock-Slater calculation results are marked with
      blue asterisk and taken from \change{Ref.~}\cite{YEH19851}. See
      Sec. \ref{sec:photo-ionis-cross} for details.}}
  \label{fig:PIC}
\end{figure*}

\section{Atomic scattering factors}
\label{sec:atom-scatt-fact}

Another observable characteristics that can be extracted from the
effective charge model are the atomic scattering factors. According to
their definition~\cite{hau2012high,prince_international_2006}, they
are expressed by the Fourier transforms of electronic density:
\begin{align}
  f(\vec{q}) = \int \rho(\vec{r})e^{i \vec{q} \cdot \vec{r}}
  d\vec{r}, \label{eq:15}
\end{align}
which in our approach can be calculated analytically (see
Appendix~\ref{sec:appendix-a.-details}).

The atomic scattering factors are very important for crystallography
and X-ray physics, since the crystal polarizability $\chi$ as the
function of X-ray frequency~$\omega_{\mathrm{r}}$, can be evaluated by
employing the following relation \cite{ahmadi_parametric_2013,
  baryshevsky2005parametric}
\begin{align}
  \chi(\vec g, \omega_{\mathrm{r}}) = \frac{4 \pi S(\vec g)}{\Omega_{0}
  \omega_{\mathrm{r}}^{2}}
  \left(-\frac{e^{2}}{m}f(\vec g)\right), \label{eq:17}
\end{align}
where $\Omega_0$ is the volume of a crystal cell, $\vec g$ the
reciprocal lattice vector, $S(\vec g)$ the structure factor of the
crystal, $m$ the electron mass and $e$ the electron charge \no{(in
  this relation $\omega_{\mathrm{r}}$ and $m$ are measured in
  $\mathrm{cm}^{-1}$, and $\Omega_{0}$ in $\mathrm{cm}^{3}$)}.

\change{In Fig.~\ref{fig:ASF} we present the results for neutral
  xenon~Xe$^{54}$, gold Au$^{79}$, lead Pb$^{82}$ and uranium U$^{92}$
  atoms. Our analytical expressions for atomic scattering factors are
  comparable to the DHF calculation to within~$25\%$.}

As a realistic example, we also evaluated the relative difference for
the emission intensity of parametric X-ray radiation (PXR)
\cite{feranchuk_theoretical_1985, baryshevsky2005parametric,
  skoromnik_radical_2017} from relativistic electrons in a
\no{tungsten W$^{74}$} crystal. The intensity of PXR is proportional
to the square of the absolute value of crystal
polarizability. Therefore, the relative difference for the intensity
of radiation with crystal polarizabilities from DHF calculation and
from effective charge model is given by
$(|\chi^{(0)}|^{2} -
|\chi_{\mathrm{DHF}}|^{2})/|\chi_{\mathrm{DHF}}|^{2}$. Consequently,
for the relevant range of frequencies and Bragg's angles the relative
accuracy was $\sim 20\%$.

\change{\section{Photoionization cross-section}}
\label{sec:photo-ionis-cross}	

\change{ As one more practical application of the effective charge
  model, we present the calculation of the total cross section for the
  photoionization of a multi-electron atom. From first principles it
  can be shown~\cite{PhysRev.134.A898} that within the dipole
  approximation, the differential cross-section for a photon with
  momentum $k$ to overcome an ionization energy $E_0$ and produce an
  outgoing electron with momentum $p$ can be calculated according to
  \cite{mikhailov1969relativistic}:
  \begin{equation}\label{PICformula}
    \frac{d \sigma}{d \Omega} = \frac{\alpha p (k+E_0)}{2 \pi k}
    \left|\int \psi^{\change{\dagger}}_f(\vec
      r)\vec{\alpha}\psi_i(\vec r) d\vec{r}\right|^2,
  \end{equation}
  where $\psi_i$ is the initial bound-state wavefunction and the final
  wavefunction can be described, as:
  \begin{equation}
    \psi_f(\vec r) = 4\pi \sum_{\kappa',m'}
    \xi_{\kappa',m'}\left(\frac{\vec{p}}{p}\right)e^{-i
      \delta_{\kappa'}}\psi_{\rm\change{\rm free}}(\vec r),
  \end{equation}
  where $\xi$ are normalized spinors \cite{lifshitz1974relativistic},
  describing the angular distribution and polarization of the outgoing
  electrons and $\delta_{\kappa'}$ are phase shifts, ensuring outgoing
  solutions.
}

\change{ Within the effective charge model, $\psi_i$ and
  $\psi_{\rm{free}}$ are described by negative and positive-energy
  solutions to the Dirac equation for hydrogen with relevant effective
  charge. Summing over all electrons in a given atom or ion, allows us
  then to analytically calculate the total effective cross-section for
  the photoionization process within the zeroth-order effective charge
  model (see Appendix B for details).  }

\change{ In the presented calculation both initial and final
  wavefunctions have been described using the value of effective
  charge found in Sec.~\ref{sec:zeroth-order-appr}, so that they
  correspond to the same effective potential. It is possible because
  the total cross-section includes the summation over a full set of
  intermediate electron excitations. The same approximation \CHK{was
    employed} in \change{Ref.~}\cite{YEH19851}.  On the other hand,
  the ionization energies $E_k$ are calculated separately for each
  electron, by finding the "valence effective charge" $Z^*_k$, defined
  by requiring the first order correction to $E_k$ to vanish. Hence it
  can be found by solving:
  \begin{equation}
    \Delta E^{(1)}_{\lambda_0}(Z^*_k)-\Delta E^{(1)}_{\lambda_k}(Z^*_k)=0,
  \end{equation}
  where $\lambda_0$ is the ground state configuration and $\lambda_k$
  the final configuration, i. e. without the ionised electron.
  Fig.~\ref{fig:PIC} presents the comparison of the results for two
  example atoms with the analogous non-relativistic calculation, as
  well as results of the Hartree-Fock-Slater calculations
  \cite{YEH19851}. It can be seen that shifting to a relativistic
  description improves the accuracy of such calculation. The results
  show that the effective charge model \change{give reasonable
    qualitative description and} can be \change{used for}
  approximating physical characteristics dependent on transition
  matrix elements.
}

\setlength{\tabcolsep}{10pt}
\setlength\LTcapwidth{0.85\textwidth}
\begin{longtable*}{|c|c|c|c|c|c|c|}
  \hline 
  Configuration, $J^{\pm}$ & $Z^{*}$ & $E^{(0)}$ & $E_{\mathrm{DHF}}$
  & $\Delta E/E_{\mathrm{DHF}} \cdot 100\%$ & CI - DFS \\
  \hline
  $1s^2,J=0^+$ & 91.7130 & -9651.35 & -9651.39 & 0.0004$\%$&-9651.45\\
  $1s^1 2s^1,J=1^+$ & 91.8623 &-6096.96 & -6097.01 & 0.0008$\%$&-6097.01\\
  $1s^1 2s^1,J=0^+$ & 91.8404 &-6093.51 & -6093.41 & -0.0016$\%$&-6097.01\\
  $1s^1 2p^1_{1/2},J=0^-$ & 91.8148  & -6089.50  & -6090.16 &0.0108$\%$ &
  -6090.17\\
  $1s^1 2p^1_{1/2},J=1^-$ & 91.8113 & -6088.95 & -6089.10 & 0.0025$\%$ &
  -6089.11\\
  $1s^1 2p^1_{3/2},J=2^-$ & 91.8487 & -5928.36  & -5928.44 & 0.0013$\%$ &
  -5928.44\\
  $1s^1 2p^1_{3/2},J=1^-$ & 91.8395 & -5927.00  & -5926.68 &-0.0054$\%$ &
  -5928.44\\
  $1s^1 3s^1,J=1^+$ & 91.9086 & -5389.83  & -5389.88 & 0.0009$\%$ &
  -5389.88\\
  $1s^1 3s^1,J=0^+$ & 91.9177  & -5388.95 & -5388.73 & -0.0041$\%$ &
  -5389.88\\
  \hline
  $1s^2 2s^1,J=\frac{1}{2}^+$ & 91.5805 &-10862.2 & -10862.4 &
  0.0018$\%$&-10862.5\\
  $1s^2 2p^1_{1/2},J=\frac{1}{2}^-$ & 91.5378 & -10850.3 & -10849.1 &
  -0.0104$\%$&-10850.8\\
  $1s^2 2p^1_{3/2},J=\frac{3}{2}^-$ & 91.5685 & -10694.8 & -10693.9 &
  -0.0084$\%$&-10695.2\\
  $1s^2 3s^1,J=\frac{1}{2}^+$ & 91.6447 & -10168.7 & -10168.9 &
  0.0020$\%$&-10169.0\\
  $1s^2 3p^1_{1/2},J=\frac{1}{2}^-$ & 91.6323 & -10165.5 & -10165.4 &
  -0.0015$\%$ &-10165.8\\
  $1s^2 3p^1_{3/2},J=\frac{3}{2}^-$ & 91.6429 & -10119.2 & -10119.0 &
  0.0004$\%$&-10119.4 \\
  $1s^2 3d^1_{3/2},J=\frac{3}{2}^+$ & 91.6399 & -10118.4 & -10118.6 &
  -0.0020$\%$&-10118.7\\
  $1s^2 3d^1_{5/2},J=\frac{5}{2}^+$ & 91.6435 & -10106.7 & 10106.9 &
  0.0022$\%$ &-10107.0\\
  \hline
  $1s^2 2s^2 2p^1_{1/2},J=\frac{1}{2}^-$ & 91.2080 & -13221.1 &
  -13222.5 & 0.0104$\%$&-13222.7\\
  $1s^2 2s^1 2p^2_{1/2},J=\frac{1}{2}^+$ & 91.1604 & -13204.8 &
  -13206.7 & 0.0140$\%$&-13206.9\\
  $1s^2 2s^2 2p^1_{3/2},J=\frac{3}{2}^-$ & 91.2339 & -13068.9 &
  -13070.2 & 0.0099$\%$&-13070.6\\
  $1s^2 2s^1 2p^1_{1/2} 2p^1_{3/2},J=\frac{3}{2}$ & 91.1972 & -13056.6 &
  -13058.8 & 0.0168$\%$&-13058.9\\
  $1s^2 2s^1 2p^1_{1/2} 2p^1_{3/2},J=\frac{5}{2}$ & 91.1914 & -13054.7
  & -13056.3 & 0.0128$\%$&-13056.5\\
  $1s^2 2p^2_{1/2} 2p^1_{3/2},J=\frac{3}{2}^-$ & 91.1327 & -13035.1 &
  -13037.2 & 0.0159$\%$ &-13037.1\\ 
  $1s^2 2s^1 2p^2_{3/2},J=\frac{1}{2}^+$ & 91.2121 & -12900.7 &
  -12900.9 & 0.0016$\%$&-13053.4\\
  $1s^2 2s^2 3s^1,J=\frac{1}{2}^+$ & 91.3255 & -12556.2 & -12557.2 &
  0.0080$\%$&-12557.7\\
  \hline
  \caption{Comparison of the first few excited energies of He-, Li-
    and B-like uranium ions \no{U$^{90+}$, U$^{89+}$ and U$^{87+}$,
      respectively}, calculated within \no{by the relativistic
      effective charge model} the zeroth-order approximation
    ($E^{(0)}$), with the solution of DHF equations obtained from
    GRASP2k program \cite{jonsson_new_2013, DYALL1989425}
    ($E_{\mathrm{DHF}}$); \no{and the relative difference between
      these two methods in percentages
      ($\Delta E/E_{\mathrm{DHF}} \cdot 100\%$)}. The last column \no{
      represents the results obtained by the configuration interaction
      Dirac-Fock-Sturm method}~\cite{Tup2003OS}. All energy values are
    in Hartree units. \no{See Sec. \ref{sec:highly-charged-ions} for
      details.}}
  \label{tab:Ions}
\end{longtable*}

\section{Conclusions and Outlook}
\label{sec:conclusions}

We have demonstrated that the relativistic effective charge model
describes multi-electron atoms and ions and provides analytic
expressions for energies, electronic densities and scattering factors
with accuracy comparable to the numerical solutions of
Dirac-Hartree-Fock equations. We have also shown that our approach is
suitable for the description of arbitrary excited states and provides
accuracy independent of the number of electrons already in the
zeroth-order approximation. Furthermore, the zeroth-order
approximation can be easily modified to include interactions with
external fields.

\no{We would like to stress, that} the introduction of the effective
charge $Z^{*}$, instead of the usage of the nuclear charge $Z$,
i.e. $Z^{*}\neq Z$, \no{is exactly the key idea,} that significantly
increases the accuracy of the zeroth-order approximation, while
rendering the complexity of all calculations low.

However, the fully relativistic description becomes much more
\CHK{complex} due to the structure of Dirac equation. First, the Dirac
equation possesses negative energy states and, therefore, there exist
matrix elements between electronic states and negative energy
ones. This significantly increases the number of required matrix
elements to be taken into account. Second, due to the mass term of the
Dirac equation the energy of the system does not have a universal
behavior of the quadratic function of the effective charge as in the
nonrelativistic case. Therefore, it is impossible to separate
explicitly the dependence of the matrix elements on the effective
charge. For this reason, before investigating the second-order
correction to the energy of the system we focused on the zeroth-order
approximation first.

At the same time, we point out that since the Dirac-Coulomb Green's
function is known in analytical closed-form \cite{wong_dirac_1985,
  swainson_unified_1991}, it is still possible to express all sums
over intermediate states \CHK{in terms of known functions} and
therefore to perform calculations of higher-order perturbative
corrections.

\change{Finally, we would like to emphasize that even though} \CHK{our
  analytical expressions are sometimes complex}, \change{all special
  functions in} \CHK{the} \change{presented calculations reduce to
  expressions containing Gamma functions and/or elementary functions
  only. Therefore all relevant evaluations of energies, electron
  densities and scattering factors can be performed without any
  numerical or convergence issues.}

\begin{acknowledgments}            
  ODS is grateful to A. Leonau, S. Bragin and U. Sinha for helpful
  discussions. This article comprises parts of the PhD thesis work of
  Kamil Dzikowski to be submitted to the Heidelberg University,
  Germany.
\end{acknowledgments}

\appendix

\section{Details of the calculation}
\label{sec:appendix-a.-details}

Here, we present all formulas needed for the fully analytical
evaluation of all integrals encountered in the calculation of the
first-order correction to the energy of the system, as well as the
computation of atomic scattering factors and electronic densities.

\no{First of all,} both the basis wave functions and all appearing
potentials can be split into their radial and angular components by
simple expansion in spherical harmonics. Thus we \CHK{obtain} the
hydrogen-like Dirac wave functions as \cite{flugge_practical_1971}
\begin{align}
  \Psi_{n_{r}ljm}
  &(\vec{r},Z^*) \nonumber
  \\
  &=
    \frac{1}{r} \left(
    \begin{array}{c}
      G_{n_{r},\kappa}(r,Z^*)\Omega_{\kappa,m}(\theta,\phi)
      \\
      F_{n_{r},\kappa}(r,Z^*)\Omega_{-\kappa,m}(\theta,\phi)
    \end{array}
  \right), \label{eq:28}
\end{align} 
with the angular part given by spherical harmonics
\begin{align}
  \Omega_{\kappa,m}
  &(\theta,\phi) \nonumber
  \\
  &= \left(
    \begin{array}{c}
      \sqrt{\frac{1}{2} - \frac{m}{2\kappa+1}}
      Y_\kappa^{m-1/2}(\theta,\phi)
      \\
      -\sqrt{\frac{1}{2} + \frac{m}{2\kappa+1}}
      Y_\kappa^{m+1/2}(\theta,\phi)
    \end{array}
  \right), \label{eq:29}
\end{align}
and the radial part by the Whittaker functions $M_{\kappa,\mu}(z)$ of
the first kind \cite{whittaker_expression_1903}
\begin{widetext}
  \begin{align}
    \begin{pmatrix}
      G_{n_r,\kappa}(Z^{*}, r)
      \\
      F_{n_r,\kappa}(Z^{*}, r)
    \end{pmatrix} 
    = N\Bigg\{\frac{n_r}{2\gamma}\left(
    \begin{array}{c}
      Z^*\alpha
      \\
      i (\gamma-\kappa)
    \end{array}\right) M_{\gamma+n_r,\gamma+\frac{1}{2}}(2 \chi  Z^*r) 
    -\rho(1+2\gamma) \left(
    \begin{array}{c}
      \frac{\gamma-\kappa}{Z^* \alpha}
      \\
      i\end{array}
    \right) M_{\gamma+n_r,\gamma-\frac{1}{2}}(2 \chi
    Z^*r)\Bigg\}, \label{eq:30}
  \end{align}
\end{widetext}
where
\begin{align*}
  \rho
  &=\frac{1}{n_r+2\gamma} \left(\kappa(n_r+\gamma) -
    \frac{\gamma}{\chi}\right),
  \\
  \chi
  &=(n_r^2+\kappa^2+2n_r\gamma)^{-\frac{1}{2}},
  \\
  \gamma
  &= \sqrt{\kappa^2-(Z^* \alpha)^2},
  \\
  \no{n_r}
  &\no{= n-|\kappa|,}
\end{align*}
\no{$n$ is the principal quantum number,} and $N$ the corresponding
normalization constant
\begin{align*}
  N=\frac{\chi}{\Gamma(2\gamma+2)} \sqrt{\Gamma(2\gamma+n_r+1)
  \frac{(\kappa+\gamma)}{n_r!} \frac{\chi Z^*}{\rho}}.
\end{align*}

The \no{relativistic angular quantum number} $\kappa$ is defined as
\begin{align}
  \kappa = \left\{
  \begin{aligned}
    l, \quad \mathrm{if}\ j = l - \frac{1}{2},
    \\
    -(l+1), \quad  \mathrm{if}\ j = l + \frac{1}{2},
  \end{aligned}\right. \label{eq:31}
\end{align}
and $\Gamma(z)$ is the Gamma function. \no{In addition, we employ the
  spherical harmonics with a negative first index, which is used in
  Mathematica~\cite{Mathematica} and is defined by the following
  identity $Y_{l}^{m} = Y_{-(l+1)}^{m}$ for $l \leq -1$ and
  $(l+1) \leq m \leq -(l+1)$ for $l \leq -1$.}

\no{We also mention here that for the convenience we expressed the Dirac
  wave functions through the Whittaker functions and not through more
  commonly used hypergeometric ones
  \cite{flugge_practical_1971}. However, our definition in
  Eq.~(\ref{eq:30}) is in full agreement with the commonly used wave
  functions from \cite{flugge_practical_1971}.}

\no{With the above definitions, we can now derive the integrals A and
  B defined in Eq.~\eqref{eq:12}.} By noting that
\begin{equation*}
  \int M_{a+\gamma,\gamma-1/2}(r) M_{a+\gamma,\gamma-1/2}(r)
  \frac{dr}{r} = \frac{\Gamma(2\gamma)^2 a!}{\Gamma(a-2\gamma)},
\end{equation*}
we get (after lengthy but straightforward simplifications)
\begin{align}
  A_{n_r \kappa}(Z^*) =
  Z^*\chi ^3\left(n_r+\frac{\kappa^2}{\gamma}\right). \label{eq:32}
\end{align}

Furthermore, in order to evaluate $B_{\nu_2,\nu_4}^{\nu_1,\nu_3}$, we
employ the expansion of the electron-electron interaction potential in
spherical harmonics \cite{landau1981quantum}
\begin{align*}
  \frac{1}{|\vec{r}-\vec{r}'|} = \sum_{l=0}^\infty \sum_{m=-l}^{l}
  \frac{4\pi}{2l+1} \frac{r_<^l}{r_>^{l+1}} Y_{l}^{m*}(\Omega)
  Y_{l}^{m}(\Omega')
\end{align*}
and note that the integration of three spherical harmonics yields $3j$
symbols \cite{Brink:104381}:
\begin{widetext}
  \begin{align}
    \int Y_{l_{1}}^{m_{1}}(\Omega) Y_{l_{2}}^{m_{2}}(\Omega)
    Y_{l_{3}}^{m_{3}}(\Omega) d\Omega =
    \sqrt{\frac{(2l_1+1)(2l_2+1)(2l_3+1)}{4 \pi}}
    \begin{pmatrix}
      l_1 & l_2 & l_3
      \\
      0 & 0 & 0
    \end{pmatrix}
              \begin{pmatrix}
		l_1 & l_2 & l_3
		\\
		m_1 & m_2 & m_3
              \end{pmatrix}. \label{eq:34}
  \end{align}
  
  This allows us to split \no{all the $B_{\nu_2,\nu_4}^{\nu_1,\nu_3}$
    coefficients} into their radial and angular parts as
  \begin{align}
    B_{\nu_2,\nu_4}^{\nu_1,\nu_3} =
    -\delta_{m_{j2}-m_{j1}}^{m_{j3}-m_{j4}} (-1)^{2
    m_{j1}-m_{j2}-m_{j3}} \sum_p \left( \Phi_{\nu_1,\nu_2}^p \otimes
    \Phi_{\nu_3,\nu_4}^p \right) \cdot
    \sigma_{\nu_1,\nu_2,\nu_3,\nu_4}^p, \label{eq:35}
  \end{align}
  \no{where $\otimes$ denotes the Kronecker product,} and the angular
  part is given by
  \begin{align}
    \Phi_{\nu_1,\nu_2}^p = \left(
    \begin{array}{c}
      \begin{pmatrix}
        \kappa_1 & \kappa_2 & p
        \\
        0 & 0 & 0
      \end{pmatrix}
		\left(
		\phi_{\kappa_1,m_{j1} -
                \frac{1}{2}}^{\kappa_2,m_{j2}-\frac{1}{2}}
		(p) -
		\phi_{\kappa_1,\frac{1}{2} -
                m_{j1}}^{\kappa_2,\frac{1}{2}-m_{j2}}
		(p)\right)
      \\
      \begin{pmatrix}
        -\kappa_1 & -\kappa_2 & p
        \\ 0 & 0 & 0
      \end{pmatrix}
                   \left(\phi_{-\kappa_1,m_{j1} -
                   \frac{1}{2}}^{-\kappa_2,m_{j2}-\frac{1}{2}}(p) -
                   \phi_{-\kappa_1,\frac{1}{2} -
                   m_{j1}}^{-\kappa_2,\frac{1}{2}-m_{j2}}(p)\right)
    \end{array}\right), \label{eq:36}
  \end{align}
  where
  \begin{align}
    \phi_{k_1,m_1}^{k_2,m_2}(p) = \sqrt{(k_1-m_1)(k_2-m_2)}
    \begin{pmatrix}
      \kappa_1 & \kappa_2 & p
      \\
      -m_{1} & m_{2} & m_{1}-m_{2}
    \end{pmatrix}. \label{eq:37}
  \end{align}
  
  Furthermore, the radial part is given by
  \begin{align}
    \sigma_{\nu_1,\nu_2,\nu_3,\nu_4}^p = \int
    {{G_{\nu_1}(r)
    G_{\nu_2}(r)}\choose{F_{\nu_1}(r)
    F_{\nu_2}}(r)} \otimes
    {{G_{\nu_3}(r') G_{\nu_4}(r')}
    \choose{F_{\nu_3}(r')
    F_{\nu_4}(r')}}
    \frac{r_<^p}{r_>^{p+1}}drdr', \label{eq:38}
  \end{align}
  where dependence on $Z^*$ has been omitted for
  clarity. Eq.~(\ref{eq:38}) can be calculated analytically by noting
  that the integral of the four Whittaker functions reads
  \cite{gradstejn_table_2009}
  \begin{align*}
    \int
    &M_{a_1+b_1,b_1-1/2}(q_1 r) M_{a_2+b_2,b_2-1/2}(q_2 r)
      M_{a_3+b_3,b_3-1/2}(q_3 r') M_{a_4+b_4,b_4-1/2}(q_4 r')
      \frac{r_<^l}{r_>^{l+1}} dr dr' \nonumber
    \\
    &=\sum_{i_1=0}^{a_1} \sum_{i_2=0}^{a_2} \sum_{i_3=0}^{a_3}
      \sum_{i_4=0}^{a_4} T_{\vec{a},\vec{b},\vec{q}}(\vec{i})
      \left(f_{i_1+i_2+b_1+b_2+l+1}^{i_3+i_4+b_3+b_4-l}
      \left(\frac{q_3+q_4}{2},\frac{q_1+q_2}{2}\right)
      + f_{i_3+i_4+b_3+b_4+l+1}^{i_1+i_2+b_1+b_2-l}
      \left(\frac{q_1+q_2}{2},\frac{q_3+q_4}{2}\right)\right),
  \end{align*}
  where
  \begin{align*}
    T_{\vec{a},\vec{b},\vec{q}}(\vec{i}) =
    \prod_{k=1}^4{{a_k}\choose{i_k}}
    \frac{\Gamma(2b_k)}{\Gamma(2b_k+i_k)}
    (-1)^{i_k}q_k^{b_k+i_k}
  \end{align*}
  and we have made use of
  \begin{align}
    \label{eq:f}
    f_{a}^{b}(x,y) = \int_0^\infty \int_r^\infty e^{-\lambda r -
    \lambda' r'} r^{a-1}
    {r'}^{b-1} dr' dr=\frac{\Gamma(a+b)}{a{\lambda'}^{a+b}}
    {_2F_1}\left(a, a+b, a + 1, -\frac{\lambda}{\lambda'}\right).
  \end{align}
  The bold $\vec a$,
  $\vec b$, $\vec q$ and $\vec i$ are lists of four values, i. e.,
  $\vec a = \{a_{1}, a_{2}, a_{3}, a_{4}\}$ with similar expressions
  for $\vec b$, $\vec q$ and~$\vec i$.
  
  \change{We would like to emphasize here that Whittaker functions,
    describing hydrogen-like bound states, are all elementary
    functions, and the simplification of the $f$ function defined in
    \eqref{eq:f} for the case of integer parameters has been described
    in some detail in \cite{dzikowski_generating_2019}.}
  
  \section{Details of the scattering factors calculation}
  
  \no{Here we calculate atomic scattering factors, as Fourier
    transforms of electronic density:}
  \begin{align}
    f_{n_r,\kappa}(q,Z^*) \no{= \int \rho_{n_r,\kappa} (r,Z^*) e^{i
    \vec{q} \cdot \vec{r}} d \vec{r}}. \label{eq:40}
  \end{align}
  Integrating out the angular dependence in \eqref{eq:28}, we get the
  radial density as
  \begin{align}
    r^2 \rho_{n_r,\kappa} (r,Z^*) =
    |G_{n_r,\kappa}(r,Z^*)|^2+|F_{n_r,\kappa}(r,Z^*)|^2, \label{eq:39}
  \end{align}
  
  Now, expanding Whittaker functions in a finite series in
  Eq.~(\ref{eq:30}) and using \cite{gradstejn_table_2009}
  \begin{align}
    \int e^{-\alpha r}r^{n-2} e^{i \vec{q} \cdot \vec{r}} d\vec{r} =
    4\pi \Gamma(n) \frac{\sin(n
    \tan^{-1}(\frac{q}{\alpha}))}{\sqrt{(\alpha^2+q^2)^n}},
    \label{eq:42}
  \end{align}
  we get
  \begin{align}
    f_{n_r,\kappa}(q,Z^*) = (N(2\gamma+1)\Gamma(2\gamma))^2 \left(
    2\kappa (\kappa-\gamma) n_r^2\sigma_1 + 4(\kappa-\gamma) \rho n_r
    \sigma_2 +
    \frac{2\kappa}{\kappa+\gamma}\rho^2\sigma_3\right), \label{eq:41}
  \end{align}
  where
  \begin{align*}
    \sigma_{1}
    &= \sum_{\substack{i=1\\j=1}}^{n_r} {{n_r-1}\choose{i-1}}
    {{n_r-1}\choose{j-1}}
    \frac{\Gamma(i+j+2\gamma)}{\Gamma(2\gamma+i+1)!
    \Gamma(2\gamma+j+1)!} \xi_{i,j}(q,Z^*),
    \\
    \sigma_{2}
    &= \sum_{\substack{i=1\\j=0}}^{n_r} {{n_r-1}\choose{i-1}}
    {{n_r}\choose{j}} \frac{\Gamma(i+j+2\gamma)}{\Gamma(2\gamma+i+1)!
    \Gamma(2\gamma+j)!} \xi_{i,j}(q,Z^*),
    \\
    \sigma_{3}
    &= \sum_{\substack{i=0\\j=0}}^{n_r} {{n_r}\choose{i}}
    {{n_r}\choose{j}}
    \frac{\Gamma(i+j+2\gamma)}{\Gamma(2\gamma+i)!\Gamma(2\gamma+j)!}
    \xi_{i,j}(q,Z^*),
    \\
    \xi_{i,j}(q,Z^*)
    &= \frac{(-1)^{i+j}}{q} \sin\left((i+j+2\gamma)
      \tan^{-1}\left(\frac{q}{2\chi Z^*}\right)\right) \left(\frac{2
      \chi Z^*}{\sqrt{(2\chi Z^*)^2+q^2}}\right)^{i+j+2\gamma}.
  \end{align*}

  \section{Details of the photoionization calculation}
  
  \change{Here we present the details of the calculation of
    Eq.~\eqref{PICformula}.  Within the framework of the effective
    charge model, we describe $\psi_i$ as a hydrogen-like wavefunction
    with effective charge by means of Eq.~\eqref{eq:28}, while
    \change{continous spectrum} solutions to the Dirac equation, which
    within the effective charge model are described by hydrogen-like
    free wavefunctions with energy $E$, momentum $p$ and efffective
    charge $Z^*$ \cite{PhysRev.134.A898}:
    \begin{equation}
      \psi_{\rm{free}} = \left(\begin{matrix}
          G_{p,\kappa'} (r,Z^*)~~ \Omega_{\kappa',m'}
          \\
          F_{p,\kappa'} (r,Z^*)~~\Omega_{-\kappa',m'}
        \end{matrix}\right) = \frac{1}{2\sqrt{p r^3}}
      \frac{|\Gamma(1+\gamma+i \nu)|}{\Gamma(2\gamma+1)}
      \left(\begin{matrix}\sqrt{1/E+1} \im(\Psi (r,Z^*))~~
          \Omega_{\kappa',m'}
          \\
          \sqrt{1/E-1} \re(\Psi
          (r,Z^*))~~\Omega_{-\kappa',m'} \end{matrix}\right),
    \end{equation}
    with $\nu = Z^* E/p$ and:
    \begin{equation}
      \Psi (r,Z^*) =(1+i)\sqrt{\frac{\kappa - i Z^*/p}{\gamma-i \nu}}
      e^{\pi/2 (\nu+i \gamma)} M_{1/2+i \nu,\gamma}(-2i p r).
    \end{equation}}
  
  \change{Using the orthogonality:
    \begin{equation}
      \int \xi_{\kappa,m}\xi_{\kappa',m'} d\Omega =
      \frac{1}{2}\delta_{\kappa,\kappa'}\delta_{m,m'},
    \end{equation}
    we can integrate Eq.~\eqref{PICformula}, to \CHK{obtain} the total
    photoionization cross section, as:
    \begin{equation}
      \sigma_{\mathrm{tot}} = \frac{\alpha p E}{k} 4\pi
      \sum_{\kappa',m'} \left|\vec{J}\right|^2,
    \end{equation}
    where:
    \begin{equation}
      \vec{J} = 
      \int\left(\begin{matrix}
          G_{p,\kappa'}(r,Z^*)~~
          \Omega_{\kappa',m'}
          \\
          F_{p,\kappa'}(r,Z^*)~~\Omega_{-\kappa',m'}
        \end{matrix}\right)^{\change{\dagger}}\left(\begin{matrix}
          0&&\vec{\sigma}
          \\
          \vec{\sigma}&&0
        \end{matrix}\right)\left(\begin{matrix}
          G_{n,\kappa}(r,Z^*)~~ \Omega_{\kappa,m}
          \\
          F_{n,\kappa}(r,Z^*)~~\Omega_{-\kappa,m}
        \end{matrix}\right) d\vec{r},
    \end{equation}
    with different Pauli matrices {$\sigma$}corresponding to different
    polarization directions. Directing the photon momentum along the
    {\it{z}} axis of our coordinate system
    ($\vec{e}^1, \vec{e}^2, \vec{k}$) and summing over photon
    polarization states, we get~\cite{lifshitz1974relativistic}:
    \begin{equation}
      \sum_{i,j,s} J^*_i J_j e_i^s e_j^s
      =\frac{1}{2}\left(|\vec{J}|^2-\frac{(\vec{J} \cdot
          \vec{k})(\vec{J^*} \cdot \vec{k})}{k^2}\right) =
      \frac{|J_x|^2+|J_y|^2}{2} = |J_x|^2,
    \end{equation}
    where in the last step we have exploited the symmetry of the
    remaining two directions. This means that the total
    photoionization cross-section can be calculated as:
    \begin{equation}
      \sigma_{\mathrm{tot}} = \frac{\alpha p E}{k} 4\pi
      \sum_{\kappa',m'} \left|\int\left(\begin{matrix}
            G_{p,\kappa'}(r,Z^*)~~ \Omega_{\kappa',m'}
            \\
            F_{p,\kappa'}(r,Z^*)~~\Omega_{-\kappa',m'}
          \end{matrix}\right)^{\change{\dagger}}\left(\begin{matrix}
            0&&\sigma_1
            \\
            \sigma_1&&0
          \end{matrix}\right)\left(\begin{matrix}
            G_{n,\kappa}(r,Z^*)~~
            \Omega_{\kappa,m}\\F_{n,\kappa}(r,Z^*)~~\Omega_{-\kappa,m}
          \end{matrix}\right)d\vec{r}\right|^2.
    \end{equation}}
  
  \change{Now, using the orthogonality of spherical harmonics, we can
    see that:
    \begin{equation}
      \int\Omega^{\change{\dagger}}_{\kappa',m'}\sigma_1\Omega_{\kappa,m}
      d\Omega=(\delta_{\kappa',\kappa} + \delta_{\kappa',-1-\kappa})
      (\delta_{m',m+1} C_{\kappa',m'} D_{\kappa,m} + \delta_{m',m-1}
      D_{\kappa',m'}C_{\kappa,m}),
    \end{equation}
    where $C_{\kappa,m}= \sqrt{\frac{1}{2} - \frac{m}{2\kappa+1}}$ and
    $D_{\kappa,m} = \sqrt{\frac{1}{2} + \frac{m}{2\kappa+1}}$ are
    numerical coefficients of spherical harmonics, see
    \change{Eq.~}\eqref{eq:29}. This gives us:
    \begin{align}
      &\left|\int\left(\begin{matrix}
            G_{p,\kappa'}(r,Z^*)~~ \Omega_{\kappa',m'}
            \\
            F_{p,\kappa'}(r,Z^*)~~\Omega_{-\kappa',m'}
          \end{matrix}\right)^{\change{\dagger}}\left(\begin{matrix}
            0&&\sigma_1
            \\
            \sigma_1&&0
          \end{matrix}\right)\left(\begin{matrix}
            G_{n,\kappa}(r,Z^*)~~ \Omega_{\kappa,m}
            \\
            F_{n,\kappa}(r,Z^*)~~\Omega_{-\kappa,m}
          \end{matrix}\right)d\vec{r}\right| \nonumber 
      \\
      &~~=J_{\kappa'}(\delta_{\kappa',-\kappa} +
        \delta_{\kappa',-1+\kappa}) (\delta_{\kappa'm',m+1}
        C_{\kappa',m'} D_{-\kappa,m} + \delta_{m',m-1} D_{\kappa',m'}
        C_{-\kappa,m} \nonumber)
      \\
      &~~+I_{-\kappa'}(\delta_{-\kappa'\kappa} +
        \delta_{-\kappa',-1-\kappa}) (\delta_{m',m+1} C_{-\kappa',m'}
        D_{\kappa,m} + \delta_{m',m-1} D_{-\kappa',m'} C_{\kappa,m}),
    \end{align}
    where the radial integrals \CHK{read}:
    \begin{equation}
      \left(\begin{matrix}
          I_{\kappa'}
          \\
          J_{\kappa'}
        \end{matrix}\right)=\int \left(\begin{matrix}
          G_{n,\kappa}(r,Z^*)~~ F^*_{p,\kappa'}(r,Z^*)
          \\
          F_{n,\kappa}(r,Z^*)~~G^*_{p,\kappa'}(r,Z^*)
        \end{matrix}\right) dr
    \end{equation}
    \CHK{and} can always be performed analytically within the
    effective charge model. Finally, we obtain the result as:
    \begin{align}\label{eq:sigmaTot}
      \sigma_{\mathrm{tot}} = \frac{\alpha p E}{k} 4\pi
      \left[\left|I_{-\kappa}\right|^2 A_{\kappa,\kappa} +
      \left|J_{-\kappa}\right|^2 A_{-\kappa,-\kappa} +
      2\re(I_{-\kappa}^* J_{-\kappa}B_\kappa) +
      \left|I_{\kappa+1}\right|^2 A_{-\kappa-1,\kappa} +
      \left|J_{\kappa-1} \right|^2A_{\kappa-1,-\kappa}\right],
    \end{align}
    where we have defined:
    \begin{align}
      &A_{\kappa,\kappa'} = \left|C_{\kappa,m+1}D_{\kappa',m}\right|^2
        + \left|C_{\kappa',m} D_{\kappa,m-1}\right|^2 
      \\
      &B_\kappa = C_{\kappa,m+1} D_{\kappa,m} C_{-\kappa,m+1}
        D_{-\kappa,m} + C_{\kappa,m} D_{\kappa,m-1} C_{-\kappa,m}
        D_{-\kappa,m-1}.
    \end{align}}
  
  \change{For the purpose of estimating the relevance of relativistic
    corrections, we take the low $p$ limit in \eqref{eq:sigmaTot} and
    average over the $m$ quantum number to obtain a non-relativistic
    formula:
    \begin{equation}
      \sigma_{\mathrm{tot}}=\frac{4\pi^2 \alpha z^2}{3 p \omega
        (2l+1)} \left(\frac{1}{l}\left|\int R_{n,l,z}(r)R_{p,l-1,z}(r)
          r^2dr\right|^2 + \frac{1}{l+1}\left| \int R_{n,l,z}(r)
          R_{p,l+1,z}(r)r^2dr\right|^2\right),
    \end{equation}
    or equivalently:
    \begin{equation}
      \sigma_{\mathrm{tot}}=\frac{4\pi^2 \alpha}{3 p (2l+1)}
      \left(1+\frac{E-E^0}{\omega}\right)\left(l \left|\int
          R_{n,l,z}(r) R_{p,l-1,z}(r)r^3dr\right|^2 + (l+1) \left|\int
          R_{n,l,z}(r) R_{p,l+1,z}(r)r^3dr\right|^2\right),
    \end{equation}
    where $E$ and $E^0$ are the ionization energy and the zeroth-order
    energy of the bound state wavefunction. For the case of $E=E^0$ it
    reduces to the standard formula \cite{YEH19851}.}
  
  \section{Solution of the relativistic Thomas-Fermi
    equation}
  \label{sec:solut-relat-tf}

  In this appendix we describe the solution of the relativistic TF
  equation \cite{gilvarry_relativistic_1954}. The equation written in
  atomic units reads \cite{waber_relativistic_1975}
  \begin{align}
    x^{1/2}\chi''(x) = \chi^{3/2}(x)\left(1 +
    \left(\frac{128}{9\pi^{2}}\right)^{1/3} \frac{Z^{4/3}}{c^{2}}
    \chi'(x) \left(1 - \frac{x
    \chi'(x)}{2\chi(x)}\right)\right)^{3/2}, \label{eq:18}
  \end{align}
  where $x = r / (b Z^{-1/3})$, $b = (9\pi^{2} / 128)^{1/3}$ and the
  dimensionless self-consistent potential $\chi(x)$ is related to
  the self-consistent potential of the TF model as
  $\phi(r) = Z \chi(rZ^{1/3}/b) - \phi_{0}$, with the constant
  $\phi_{0}$ defined from the normalization. For neutral atoms
  $\phi_{0}$ equals zero. In the nonrelativistic limit\no{, i.e., when the
    speed of light tends to infinity} the relativistic TF equation
  \no{reduces to} its nonrelativistic \no{counterpart}.
\end{widetext}

The TF equation must be complemented with boundary conditions, which
for neutral atoms are given by \cite{waber_relativistic_1975}
\begin{align}
  \chi(0) = 1, \quad \chi(\infty) = 0, \label{eq:19}
\end{align}
and for ions \cite{marini_relativistic_1981}
\begin{align}
  \chi(0) = 1, \quad -x_{c}\chi'(x_{c}) = 1 - N/Z, \label{eq:20}
\end{align}
respectively. Here we also employed the fact that $\chi(x_{c}) = 0$.

As was mentioned in the introduction solution of the TF equation is a
nontrivial mathematical problem since it represents a boundary value
problem on a semi-infinite interval. In order to solve the equation,
we used the shooting method. For neutral atoms we reformulated the
boundary value problem as an initial value one
\begin{align}
  \chi_{0} = 0, \quad \chi'(0) = \mu, \label{eq:21}
\end{align}
where $\mu$ represents a parameter. Consequently, we were seeking for
the root of the equation $\chi(\no{x}, \mu) = 0$, where $\no{x}$ we
changed from \no{some small value} to \no{a very large one}. For every
$\no{x}$ we were solving Eq.~(\ref{eq:21}) by varying $\mu$. With this
we obtained the following solutions
\begin{align}
  \mu_{\mathrm{Xe}}
  &= -1.50965873266, \quad \chi(80,
    \mu_{\mathrm{Xe}}) \approx 10^{-6}, \label{eq:22}
  \\
  \mu_{\mathrm{U}}
  &= -1.49103044294, \quad \chi(80,
    \mu_{\mathrm{U}}) \approx 10^{-6} \label{eq:23}
\end{align}
for \no{considered} atoms.

For ions we used a similar strategy, however, we ``shot'' from
infinity. In this case the boundary value problem is already written
as the initial value one
\begin{align}
  \chi(x_{c}) = 0, \quad \chi'(x_{c}) =
  -\frac{1-N/Z}{x_{c}}. \label{eq:24}
\end{align}

For this reason we simply varied the value of $x_{c}$ till the value
of $\chi$ at zero becomes one. With this we got
\begin{align}
  x_{c}
  &= 0.34635, \quad \chi(10^{-6}) \approx 1, \label{eq:25}
  \\
  x_{c}
  &= 0.47890, \quad \chi(10^{-6}) \approx 1. \label{eq:26}
\end{align}

Finally, the density of the atom or ion is expressed through the
self-consistent potential as
\begin{widetext}
  \begin{align}
    \rho(r) = \frac{8\sqrt{2}}{3\pi} \left(\frac{Z
    \chi(x)}{r} - \phi_{0}\right)^{3/2} \left(1 +
    \left(\frac{128}{9\pi^{2}}\right)^{1/3} \frac{Z^{4/3}}{c^{2}}
    \chi'(x) \left(1 - \frac{x
    \chi'(x)}{2\chi(x)}\right)\right)^{3/2}. \label{eq:27}
  \end{align}

  \section{Values of ground state energies of the first 100 neutral atoms}
  \label{sec:values-ground-state}
  
  \setlength{\tabcolsep}{10pt}
  \setlength\LTcapwidth{0.94\textwidth}
  \begin{longtable*}{|ccccc|ccccc|} 
    \hline \hline
    $Z$ &$Z^{*}_{\mathrm{R}}$ & $E^{(0)}_{\mathrm{NR}}$ & $E^{(0)}_{\mathrm{R}}$ & $E_{\mathrm{DHF}}$ & $Z$ & $Z^{*}_{\mathrm{R}}$ & $E^{(0)}_{\mathrm{NR}}$ & $E^{(0)}_{\mathrm{R}}$ & $E_{\mathrm{DHF}}$ \\
    \hline
    1 & 1.00000 & -0.50000 & -0.50000 & -0.50000 & 51 & 40.7062 & -5974.00 & -6160.94 & -6475.24 \\
    2 & 1.68749 & -2.84766 & -2.84772 & -2.86175 & 52 & 41.5615 & -6259.79 & -6461.59 & -6788.06 \\
    3 & 2.54539 & -7.28906 & -7.28951 & -7.43327 & 53 & 42.4156 & -6553.19 & -6770.65 & -7109.76 \\
    4 & 3.37163 & -14.2096 & -14.2121 & -14.5752 & 54 & 43.2653 & -6854.26 & -7087.18 & -7440.46 \\
    5 & 4.15118 & -23.6936 & -23.7003 & -24.5350 & 55 & 44.1573 & -7165.57 & -7415.35 & -7779.91 \\
    6 & 4.90693 & -36.2016 & -36.1296 & -37.6732 & 56 & 45.0484 & -7484.40 & -7752.08 & -8128.34 \\
    7 & 5.64987 & -52.0662 & -51.8941 & -54.3229 & 57 & 45.7984 & -7804.64 & -8083.37 & -8485.87 \\
    8 & 6.42240 & -71.2844 & -72.2209 & -74.8172 & 58 & 46.5481 & -8125.81 & -8423.60 & -8852.82 \\
    9 & 7.17595 & -94.4525 & -96.6125 & -99.4897 & 59 & 47.2966 & -8447.55 & -8772.63 & -9229.40 \\
    10 & 7.88116 & -121.908 & -124.316 & -128.674 & 60 & 48.0432 & -8783.92 & -9130.29 & -9615.86 \\
    11 & 8.72835 & -154.020 & -156.740 & -162.053 & 61 & 48.7880 & -9127.99 & -9496.65 & -10012.3 \\
    12 & 9.56796 & -190.415 & -193.471 & -199.901 & 62 & 49.5310 & -9479.96 & -9871.77 & -10418.8 \\
    13 & 10.3870 & -230.579 & -234.059 & -242.286 & 63 & 50.2713 & -9839.95 & -10255.2 & -10835.5 \\
    14 & 11.1991 & -275.254 & -279.124 & -289.403 & 64 & 51.0151 & -10216.4 & -10649.8 & -11262.6 \\
    15 & 12.0048 & -324.603 & -328.816 & -341.420 & 65 & 51.7609 & -10582.4 & -11055.1 & -11700.3 \\
    16 & 12.8193 & -378.517 & -384.172 & -398.503 & 66 & 52.5067 & -10965.9 & -11470.4 & -12148.7 \\
    17 & 13.6272 & -437.400 & -444.551 & -460.821 & 67 & 53.2510 & -11357.4 & -11895.2 & -12607.8 \\
    18 & 14.4170 & -501.418 & -509.263 & -528.540 & 68 & 53.9927 & -11757.1 & -12329.1 & -13078.0 \\
    19 & 15.2858 & -571.305 & -579.971 & -601.352 & 69 & 54.7306 & -12165.2 & -12771.4 & -13559.3 \\
    20 & 16.1505 & -646.244 & -655.816 & -679.502 & 70 & 55.4635 & -12581.8 & -13221.8 & -14051.9 \\
    21 & 16.9029 & -723.779 & -734.443 & -763.133 & 71 & 56.2925 & -13017.6 & -13695.5 & -14555.9 \\
    22 & 17.6518 & -806.609 & -818.525 & -852.531 & 72 & 57.1215 & -13462.0 & -14179.9 & -15071.3 \\
    23 & 18.3949 & -894.773 & -907.982 & -947.852 & 73 & 57.9500 & -13915.1 & -14674.8 & -15598.3 \\
    24 & 19.1329 & -984.973 & -1002.97 & -1049.21 & 74 & 58.7777 & -14376.8 & -15180.3 & -16136.9 \\
    25 & 19.8643 & -1087.71 & -1103.38 & -1156.87 & 75 & 59.6048 & -14847.3 & -15696.0 & -16687.4 \\
    26 & 20.6041 & -1192.25 & -1211.07 & -1270.88 & 76 & 60.4345 & -15326.2 & -16224.1 & -17249.9 \\
    27 & 21.3454 & -1302.72 & -1325.53 & -1391.42 & 77 & 61.2652 & -15813.9 & -16764.2 & -17824.6 \\
    28 & 22.0817 & -1419.13 & -1446.14 & -1518.64 & 78 & 62.0954 & -16300.8 & -17315.4 & -18400.7 \\
    29 & 22.8086 & -1536.57 & -1572.35 & -1652.71 & 79 & 62.9241 & -16806.4 & -17877.3 & -19011.3 \\
    30 & 23.5219 & -1670.43 & -1703.53 & -1793.78 & 80 & 63.7501 & -17330.8 & -18449.2 & -19623.5 \\
    31 & 24.3524 & -1809.22 & -1845.33 & -1941.63 & 81 & 64.6315 & -17861.7 & -19042.6 & -20248.3 \\
    32 & 25.1804 & -1954.42 & -1993.69 & -2096.42 & 82 & 65.5128 & -18401.7 & -19647.9 & -20886.0 \\
    33 & 26.0059 & -2106.13 & -2148.64 & -2258.28 & 83 & 66.3952 & -18950.8 & -20264.7 & -21536.7 \\
    34 & 26.8353 & -2264.11 & -2311.42 & -2427.30 & 84 & 67.2796 & -19508.5 & -20894.9 & -22200.7 \\
    35 & 27.6621 & -2428.77 & -2481.13 & -2603.59 & 85 & 68.1641 & -20075.4 & -21537.4 & -22878.2 \\
    36 & 28.4813 & -2600.19 & -2656.87 & -2787.28 & 86 & 69.0467 & -20651.5 & -22191.1 & -23561.1 \\
    37 & 29.3581 & -2780.21 & -2841.74 & -2978.07 & 87 & 69.9636 & -21241.1 & -22863.5 & -24237.8 \\
    38 & 30.2331 & -2966.85 & -3033.62 & -3176.18 & 88 & 70.8808 & -21839.6 & -23548.8 & -24992.3 \\
    39 & 31.0303 & -3155.02 & -3227.45 & -3381.68 & 89 & 71.6925 & -22437.9 & -24221.6 & -25724.9 \\
    40 & 31.8263 & -3350.00 & -3428.61 & -3594.81 & 90 & 72.5050 & -23045.4 & -24907.7 & -26471.9 \\
    41 & 32.6201 & -3546.33 & -3636.95 & -3815.67 & 91 & 73.3179 & -23639.5 & -25606.9 & -27233.7 \\
    42 & 33.4115 & -3755.01 & -3852.55 & -4044.45 & 92 & 74.1309 & -24254.1 & -26319.2 & -28010.5 \\
    43 & 34.2000 & -3976.23 & -4075.25 & -4281.19 & 93 & 74.9439 & -24878.0 & -27044.6 & -28802.9 \\
    44 & 34.9923 & -4192.63 & -4306.84 & -4526.11 & 94 & 75.7568 & -25499.2 & -27783.2 & -29610.8 \\
    45 & 35.7855 & -4422.14 & -4546.83 & -4779.23 & 95 & 76.5699 & -26141.7 & -28534.8 & -30434.9 \\
    46 & 36.5766 & -4652.44 & -4794.58 & -5040.71 & 96 & 77.3858 & -26805.4 & -29302.3 & -31275.1 \\
    47 & 37.3634 & -4902.97 & -5049.59 & -5310.66 & 97 & 78.2039 & -27466.1 & -30085.3 & -32132.1 \\
    48 & 38.1440 & -5160.83 & -5311.30 & -5589.05 & 98 & 79.0232 & -28124.0 & -30883.5 & -33006.0 \\
    49 & 38.9993 & -5424.43 & -5586.42 & -5875.84 & 99 & 79.8430 & -28803.8 & -31696.2 & -33897.2 \\
    50 & 39.8532 & -5695.47 & -5869.68 & -6171.21 & 100 & 80.6627 & -29493.1 & -32523.3 & -34806.3 \\
    \hline \hline
    \caption{Relativistic effective charge $Z^{*}$ and the comparison of
      the energy in a.u. (Hartree) of the zeroth-order approximation of
      the effective charge model in nonrelativistic and relativistic
      approach with the values obtained from numerical solutions of DHF
      equations \cite{DESCLAUX1973311}. Here $Z^{*}_{\mathrm{R}}$ is the
      effective charge, $E^{(0)}_{\mathrm{NR}}$ is the nonrelativistic
      zeroth-order energy, $E_{\mathrm{R}}^{(0)}$ is the relativistic
      zeroth-order energy, $E_{\mathrm{DHF}}$ is the ground state energy
      obtained via solution of DHF
      equations.}\label{tab:ground_state_energies}
  \end{longtable*}
  
\end{widetext}

\bibliography{Relativistic0}

\end{document}